\documentclass[aps,prc,showpacs,nofootinbib,twocolumn]{revtex4-1}

\usepackage{epsfig}
\usepackage{graphicx,amsmath}
\usepackage{color}
\newcommand\ba{\begin{eqnarray}}
\newcommand\ea{\end{eqnarray}}

\newcommand{\be}{\begin{equation}}
\newcommand{\ee}{\end{equation}}
\newcommand{\bas}{\begin{eqnarray*}}
\newcommand{\eas}{\end{eqnarray*}}
\begin{document}
\title{\bf \large Fractal properties applied to hadron spectroscopy}

\author{B. Tatischeff$^{1,2}$\\
$^{1}$Univ Paris-Sud, IPNO, UMR-8608, Orsay, F-91405\\
$^{2}$CNRS/IN2P3, Orsay, F-91405}
\thanks{tati@ipno.in2p3.fr}

\pacs{14.20,-c, 14.40.-n}

\vspace*{1cm}
\begin{abstract}
A contribution is presented to the study of hadron spectroscopy through the use of fractals and discrete scale invariance implying log-periodic corrections to continuous scaling.  The masses of  mesons and baryons, reported by the Particle Data Group (PDG), are properly fitted with help of the  equation derived from the discrete-scale invariance (DSI) model. The same property is observed for the mass ratios between different particle species. This is also the case for total widths of several hadronic species.

Each fitted parameter, as a function of the hadronic masses, displays the same distribution for all hadronic species. Several masses of still unobserved mesons and baryons are tentatively predicted.
\end{abstract}
\maketitle
\section{Introduction}
The large and increasing number of mesons and baryons suggests the need for a new classification, in addition to those already existing, based on their quark and gluon nature and quantum numbers (isospin, spin, charge conjugation and parity). A possible way is to look for eventuel fractal properties of these particles. Up to now, the very powerful concept of fractals \cite{mandelbrot} was applied to several fields in physics \cite{nottale1}, but not really used to study the physical masses of the particles. The present work is in continuity with two previous papers where the same ideas were studied. 

The first paper presented several relations, relying between themselves the masses of the two quark families: $m_u, m_c$, and $m_t$ in the one hand, and $m_d, m_s$, and $m_b$ in the other hand \cite{btib}. The same study, presented relations between gauge boson masses, and another relation between lepton masses \cite{btib}. These relations suggest that the particle physics masses should also follow fractal properties.

The second paper looked at the the fractal properties in coupling constants of fundamental forces, in atomic energies and in elementary particle masses \cite{boris}.

The hadron level-spacing was studied some time ago \cite{pascalutsa}.
In this work, it was shown that, when separated into multiplets characterized by sets of definite quantum numbers and different flavors, the masses are well described by the Wigner surmise for $\beta$ = 1. Our discussion considers consecutive hadron level masses, rather than level mass spacings, then considers consecutive hadron level mass ratios, separately for different species, but not separately for different quantum numbers. 

The hadron spectrum was recently analysed within the "Chaos in Quantum Chronodynamics" model \cite{goldfain}. It was shown that the meson and baryon spectra "obey a scaling hierarchy with critical exponents ordered in natural progression". This study applies to the fundamental masses of different species of mesons and baryons and not to the mass ratios inside each species.

The hadron masses were recently computed within the Anti-de Sitter space, conformal field theory \cite{brodsky}. It was found that "the predicted mass spectrum in the truncated space model is linear M~$\propto$~LO, at high orbital angular momentum".
\section{About the scale invariance model}
\subsection{The fractal characteristic of hadronic masses}
The fractal concept stipulates that the same physical laws apply for different scales of the given physics. We summarize here very briefly the concept of continuous and discrete scale invariances transcribing the developements of D. Sornette \cite{sornette} and L. Nottale \cite{nottale}, as  already reminded in \cite{boris}.
The concept of {\it continuous} scale invariance is defined by the following way: an observable O(x), depending on the variable x,  is scale invariant under the arbitrary change x $\to~\lambda$x, if there is a number $\mu(\lambda)$ such that  $\mu$ = O(x) / O($\lambda$x). $\lambda$ is the fundamental scaling ratio. The solution of  O(x) is:
\be
 O(x) = C x ^{\alpha}, 
\ee 
 where  $\alpha = -ln\mu/ln\lambda$.\\

The relative value of the observable, at two different scales, depends only on $\mu$, the ratio of the two scales O($\lambda$x)/O(x) and does not depend on x. We have therefore " a continuous translational invariance expressed on the logarithms of the variables".
If the distribution of the logarithm of O(x) versus the logarithm of x, displays a straight line, we expect that a relation exists allowing the calculation of the masses, starting from the lowest mass.
\subsection{The discrete scale invariance characteristic of hadronic masses}
 The {\it discrete} scale invariance (DSI), in opposition to the 
 {\it continuous} scale invariance, is observed when the scale invariance is only observed for specific choices of $\lambda$. 
Its signature is the presence of power laws with complex exponents $\alpha$ inducing log-periodic corrections to scaling \cite{sornette}. In case of DSI, the $\alpha$ exponent is now \\
\be
\hspace*{8.mm}\alpha = -ln\mu/ln~\lambda + i 2n\pi/ln\lambda \\
\ee
where n is an arbitrary integer. The {\it continuous} scale invariance is obtained for the special case n = 0, then $\alpha$ becomes real.

The critical exponent "s" is defined by $\mu$ = $\lambda^{s}$. Defining $\Omega$ = 1/ln$\lambda$, we obtain $\alpha$ = -s + i 2$\pi \Omega$. The most general form of distributions following DSI was given by Sornette \cite{sornette}. We apply it to the ratio of two hadronic adjacent masses of a given species, "f(r)", where "r" is the rank of the distribution.
Following the most general form of the mass ratio distribution f(r):\\
\be
 f(r) = C\hspace*{1.mm}(|r-r_{c}|)^{s}\hspace*{1.mm}[1 + a_{1}\hspace*{1.mm}cos(2\pi\hspace*{1.mm}\Omega\hspace*{1.mm}
ln\hspace*{1.mm}(|r-r_{c}|) + \Psi)]\\
\ee
where we have omitted the imaginary part of f(r). 
 
 C is a normalization constant. $a_{1}$ measures the amplitude of the log-periodic correction to continuous scaling, and $\Psi$ is a phase in the cosine.
"$r_{c}$" is the  critical rank, which describes the transition from one phase to another. It is underdetermined, but widely larger than the experimental "r" values. This is the general situation of hadronic mass ratios.
Such undetermination has very few consequences on all parameters, except
on $\Omega$. When increasing "$r_{c}$" from 30 to 40, $\Omega$ increases approximately by a factor 1.36. Therefore "$r_{c}$"  was arbitrarily fixed to 
"$r_{c}$" = 40 for all following studies, allowing to look at all parameter variations.

In summary, the signature of scale invariances is the existence of power laws. The exponent $\alpha$ is real if we have continuous scale invariances  and is complex in case of discrete scale invariances, and then gives rise to log-periodic corrections.

 We first plot the log of the studied quantity, versus the log of its rank (ln(R)), defined as being the mass number from the lighter to the heavier studied mass.
All masses are expressed in MeV.
  Then, when the statistics will allow it, we study the ratio of successive values fitted by equation (3). 
\section{PDG Meson Masses}
All mesonic masses, except those specifically indicated, are taken from the review of the Particle Data Group (PDG) \cite{pdg}.
Many mesons are not firmly established, even if many of them are given in PDG with their quantum numbers ${\it I(J^{P})}$. These masses, omitted from the PDG summary table, are however kept in 
our study, which therefore will have to be improved after new meson mass determinations (observations, confirmations or eliminations). Then new hadronic
masses, tentatively observed up to january 2011, were introduced in this study. The masses, corresponding to large "r" for all species, are clearly uncertain, and will have to be improved with time. 

\subsection{Light unflavored PDG meson masses}
Many unflavored mesonic masses, only quoted in PDG inside the "Other Light Mesons" section \cite{pdg}, were observed at LEAR (CERN) using the 
$p - {\bar p}$ annihilation measurements \cite{anisovitch} (the Crystal Barrel data). We introduce these mesons, using their reliability discused in \cite{bugg}. 
Figure 1 shows the log-log plot of the PDG light unflavored mesons.  We observe a staircase shape. Here only the PDG data are introduced, in order to increase the low mass range and therefore increase the staircase shape.
\begin{figure}[ht]
\caption{Log-log plot of  PDG unflavored meson masses (MeV).} 
\hspace*{-3.mm}
\scalebox{1}[1.5]{
\includegraphics[bb=14 220 530 558,clip,scale=0.5]{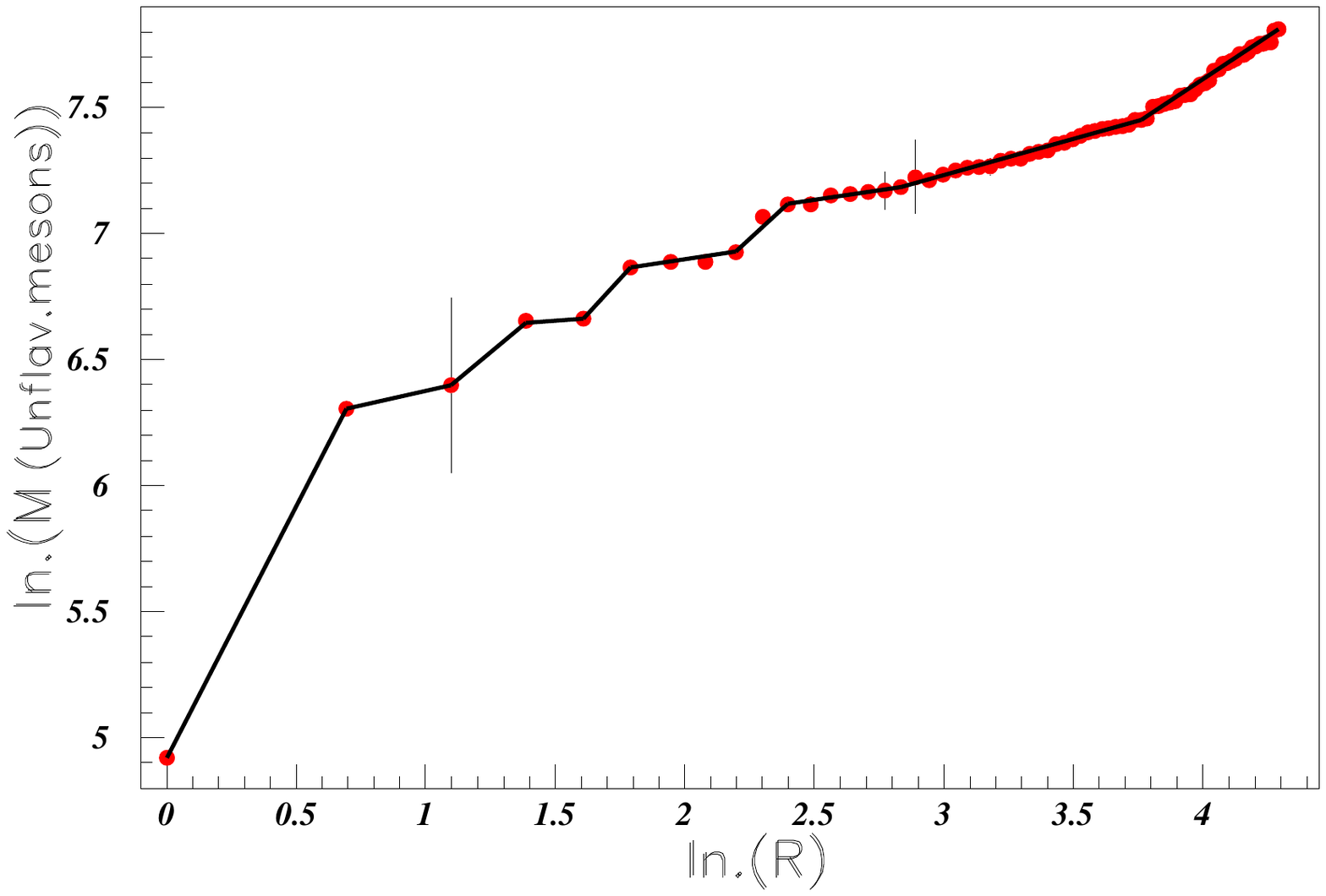}}
\end{figure}
We observe a jagged shape for the first six masses, which suggests  a possible double fractal property. We observe also that the product of the P parity by the G parity, is successively even and odd for the first thirteen unflavored meson masses, up to {\it f$_{2}$(1270)}. 

Therefore we plot for all unflavored mesons two separated log-log distributions in figure~2,  the even and odd PG parity masses being considered separately.
\begin{figure}[ht]
\begin{center}
\caption{Log-log plot of unflavored meson masses (MeV), separated as even or odd PG parity products. Full circles (red on line) are for even PG parity masses; full stars (blue on line), are for odd PG parity masses.}
\hspace*{-3.mm}
\scalebox{1}[1.5]{
\includegraphics[bb=14 230 530 558,clip,scale=0.5]{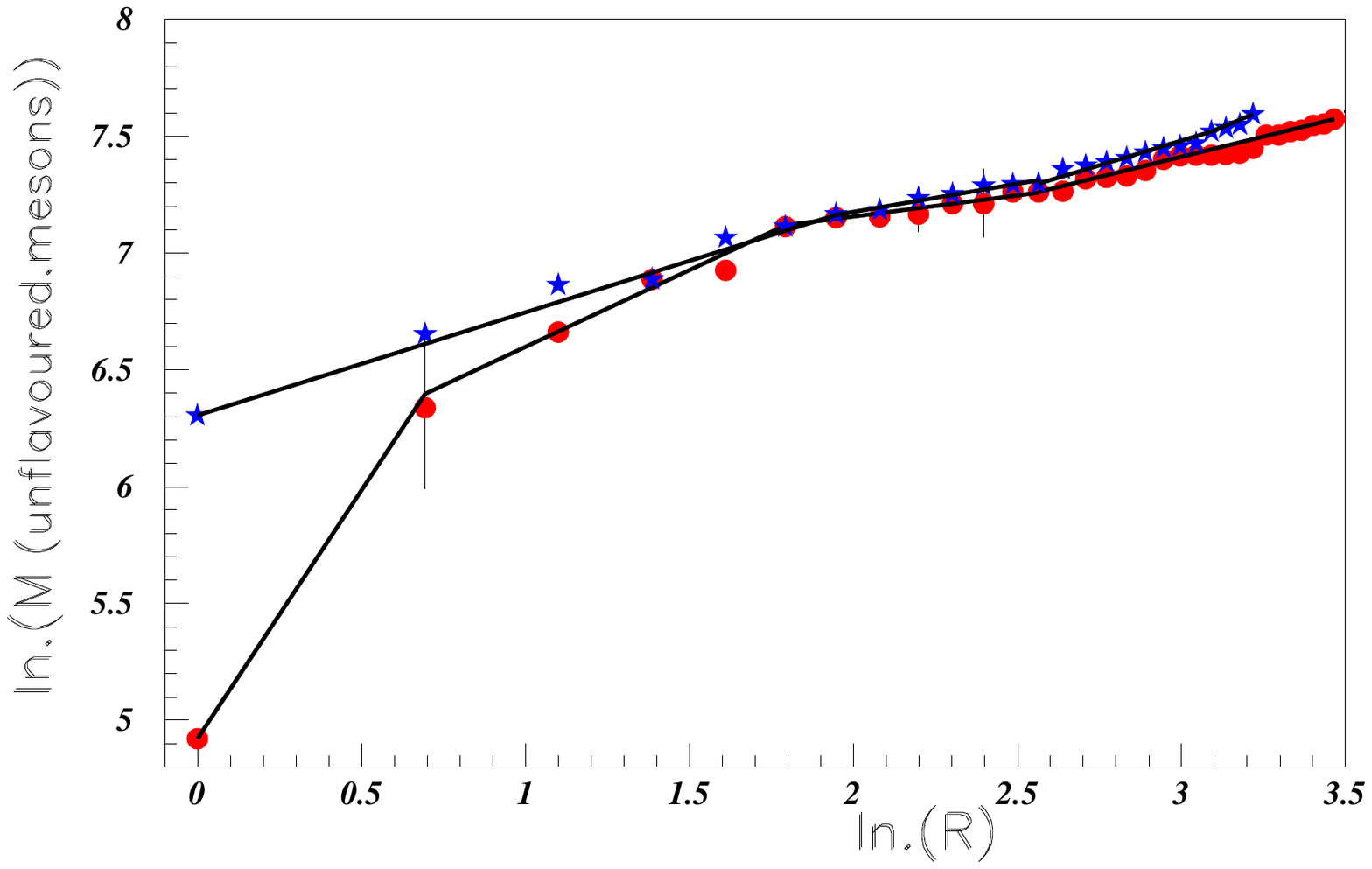}}
\end{center}
\end{figure}

The staircase shape is removed now. 
  The figure shows a few number of small steps for masses larger than 
  M~=~1474~MeV (rank 13 in this PG parity product separated plot), suggesting a possible lack of some  mesons. We deduce the presence of power-law sequences, for both ¨even¨ and ¨odd¨ PG parities product. These  relations could eventually help to predict the masses of unflavored mesons, built on the pion (or the $\eta$) mass. 

The $m_{r+1}/m_{r}$ unflavored mesonic mass ratio is plotted in figure~3. 
\begin{figure}[ht]
\begin{center}
\caption{Ratios of $m_{r+1}/m_{r}$ masses, for unflavoured mesons. Insert (a) shows the first twenty ratios, insert (b) shows the ratios between rank 10 and 50.}
\hspace*{-3.mm}
\scalebox{1}[1.5]{
\includegraphics[bb=14 137 520 550,clip,scale=0.5]{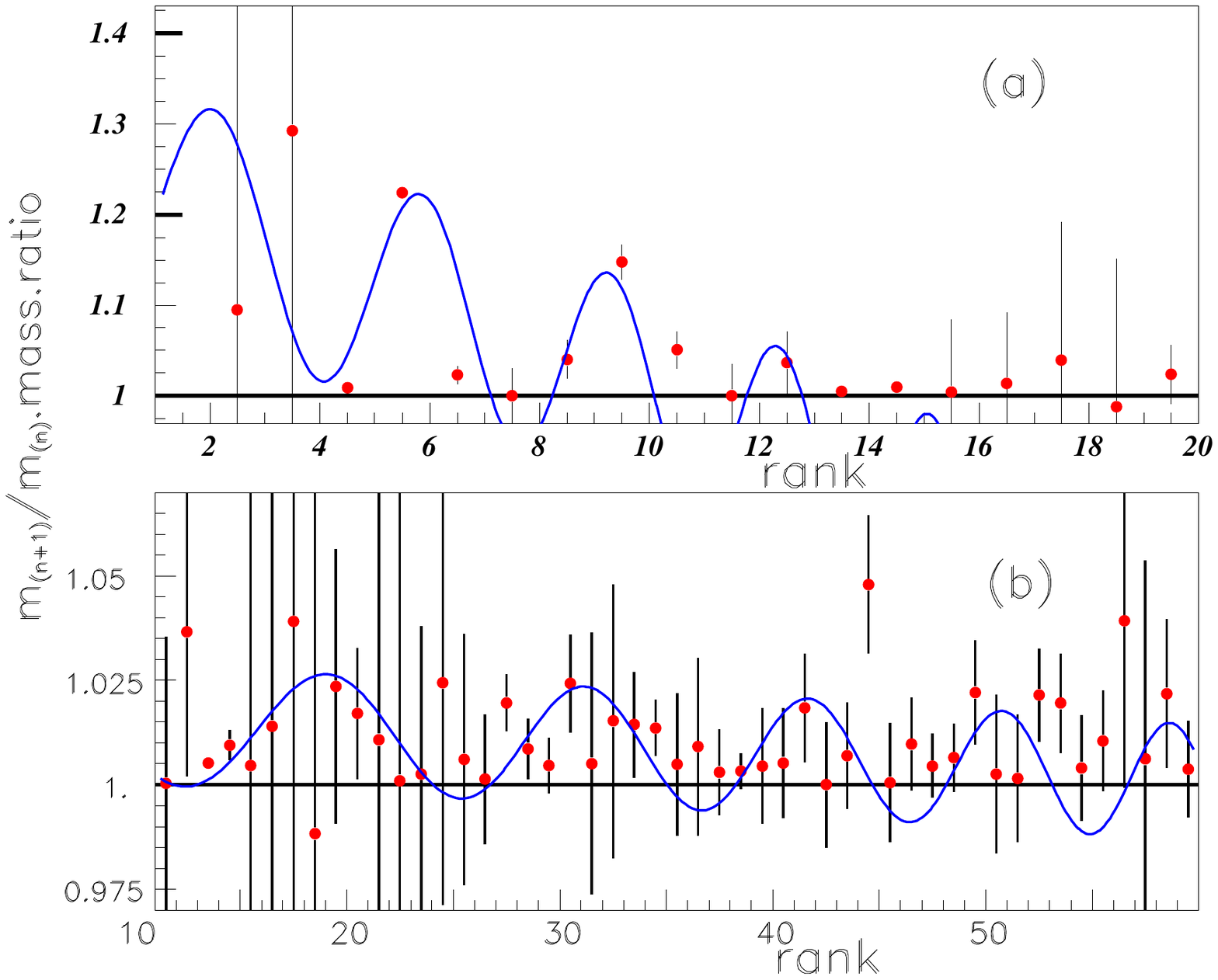}}
\end{center}
\end{figure}
The data are presented using two inserts. Insert (a) shows the low rank ratios and fit peformed using equation (3). Insert (b) shows the ratio in a larger rank range. In this range, the good alignement of the log-log distribution, shown in figure~2, could suggest a better justification of the analytical fit with an unique set of parameters. However the error bars are so large that the comparison between data and fit is almost valueless.
It is clear that this species contains a large number of unprecised masses.

 The first ratio is much larger than the other ratio values, and is  removed from the figure. It corresponds to the large mass difference between the two first pions (see figure~2). The masses of two {\it $f_{0}$} mesons are very imprecise. We take an unprecision for them to be $\Delta$M~=~200~MeV, therefore both masses are therefore M~=~600~$\pm$~200~MeV and 1370~$\pm$~200~MeV. 
 
 The masses are introduced in an increasing order, therefore, in a few cases, the order is not exactly the same as given by PDG.  Each ratio is plotted at the mean absisse between the absissa of the '' r '' and  the '' r+1 ''  masses.
\subsection{Strange meson masses}
The log-log distribution of strange meson masses is shown in figure~4. 
\begin{figure}[ht]
\begin{center}
\caption{Log-log plot of strange meson masses (MeV).}
\hspace*{-3.mm}
\vspace*{5.mm}
\scalebox{1}[0.7]{
\includegraphics[bb=14 230 530 558,clip,scale=0.5]{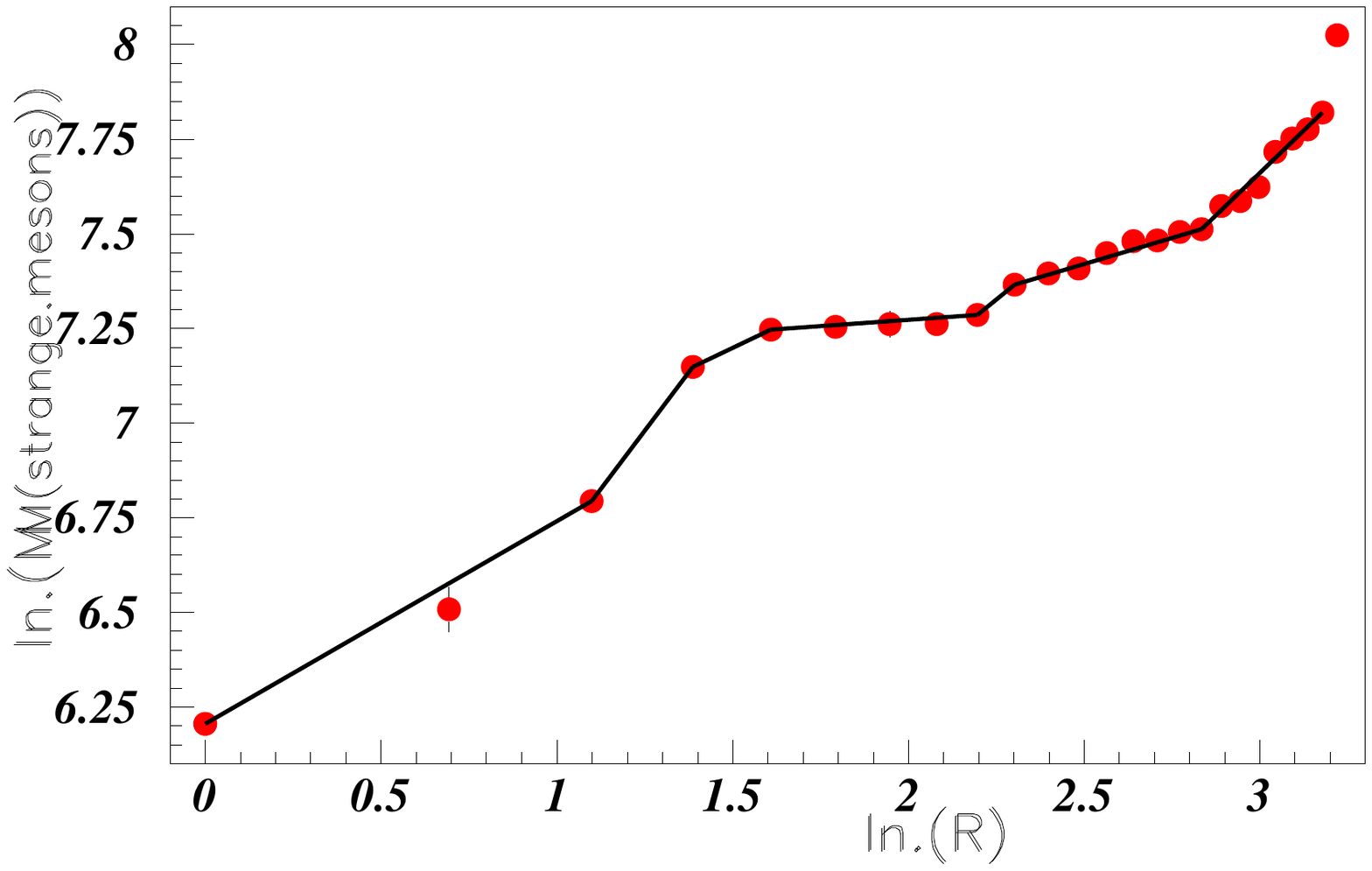}}
\end{center}
\end{figure}
The figure suggests a possible underestimation of the $K^{*}_{0}$(800) or $\kappa$ mass (the second data), reported as m = 672 $\pm$40~MeV \cite{pdg}.

 Figure~5 shows the ratio of successive strange meson masses.
\begin{figure}[ht]
\begin{center}
\caption{Ratios of $m_{r+1}/m_{r}$ masses, for strange meson masses.}
\hspace*{-3.mm}
\includegraphics[bb=9 220 530 558,clip,scale=0.5]{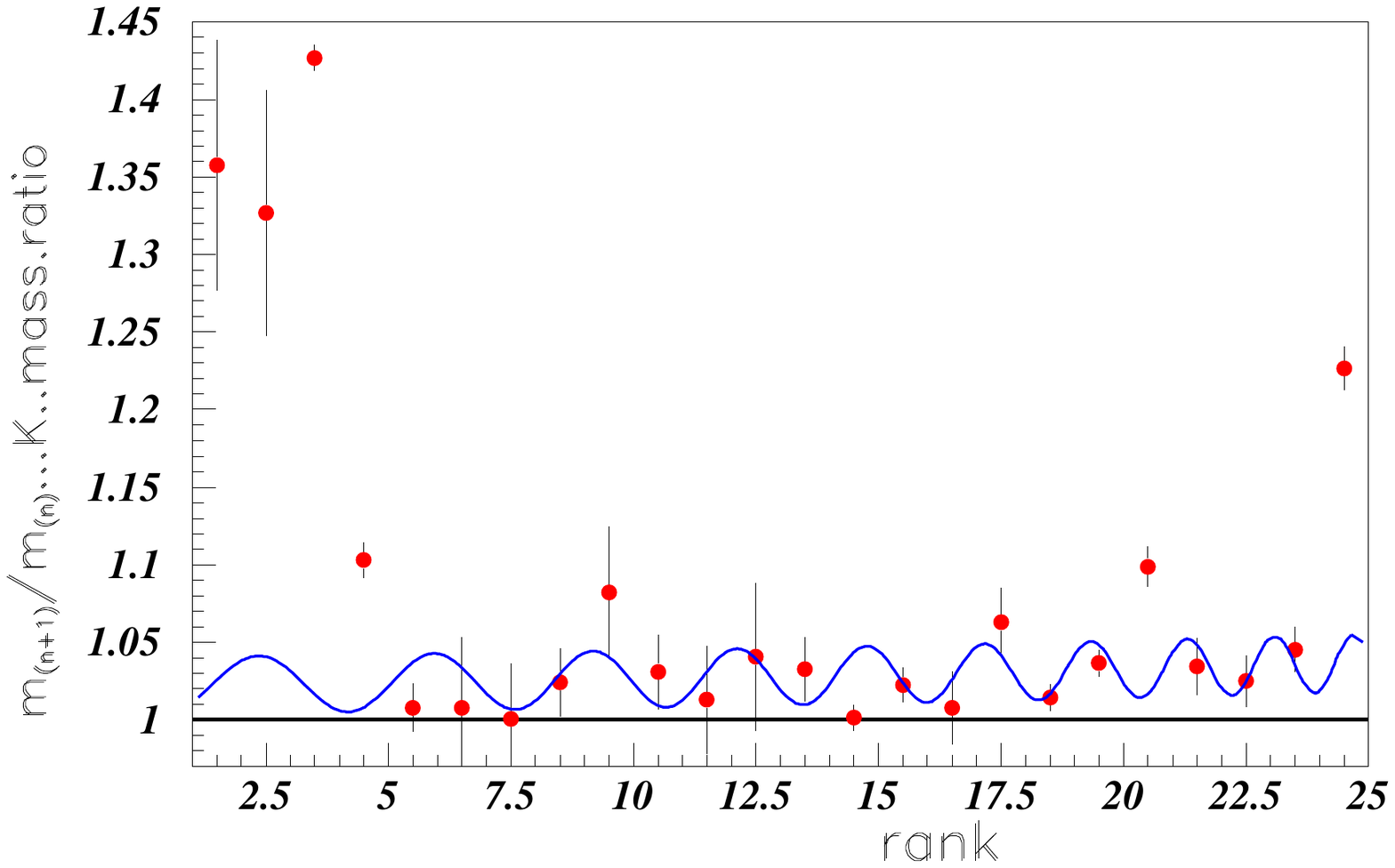}
\end{center}
\end{figure}
We observe, here again, similar behaviour as the one noted previously for the unflavoured mesons, namely an important peak at low  "r" followed by oscillations
blurred by relatively large error bars. The curve shows the result of the fit performed using equation (3). Here again, the experimental peak at low  "r" is outside the fit.
\subsection{Charmed meson masses}
The log-log distribution of the charmed meson masses is plotted in figure~6. 
 In addition to the charmed meson masses reported by PDG, several masses have been observed by the BaBar Collaboration \cite{babar} \cite{babar1} at M = 2539.4, 2608.7,  2752.4,  2710,  2763.3, and 2862~MeV. They are introduced in the figures.  Their strong decay have been analysed by \cite{zhong} \cite{wang} \cite{li}.
\begin{figure}[ht]
\begin{center}
\caption{Log-log plot of charmed meson masses (MeV).}
\hspace*{-3.mm}
\scalebox{1}[0.7]{
\includegraphics[bb=14 230 530 558,clip,scale=0.5]{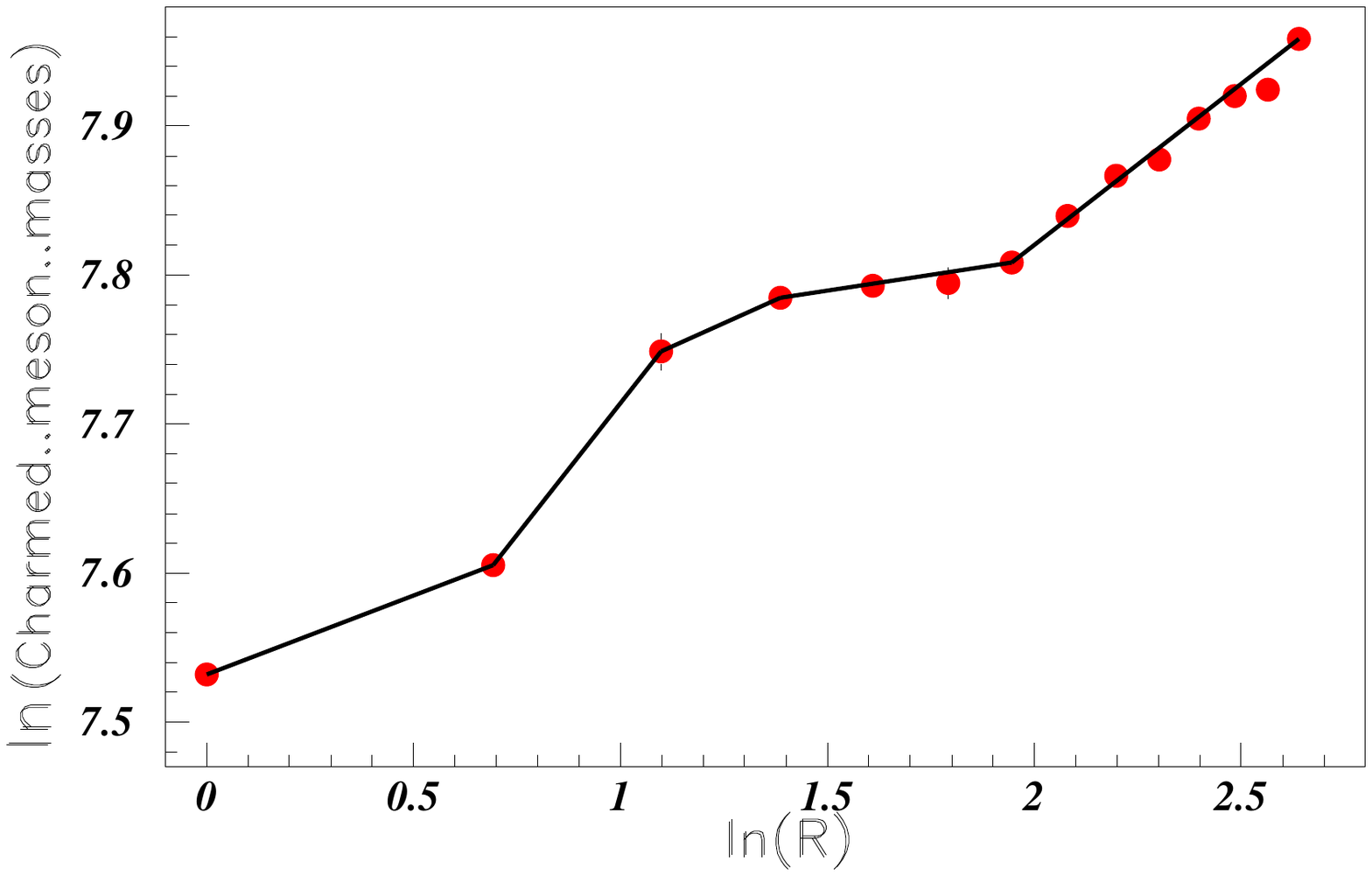}}
\end{center}
\end{figure}
 Figure~7 shows the ratio of successive charmed meson masses. 
\begin{figure}[ht]
\begin{center}
\caption{Ratios of $m_{r+1}/m_{r}$ masses, for charmed meson masses.}
\hspace*{-3.mm}
\includegraphics[bb=9 220 530 558,clip,scale=0.5]{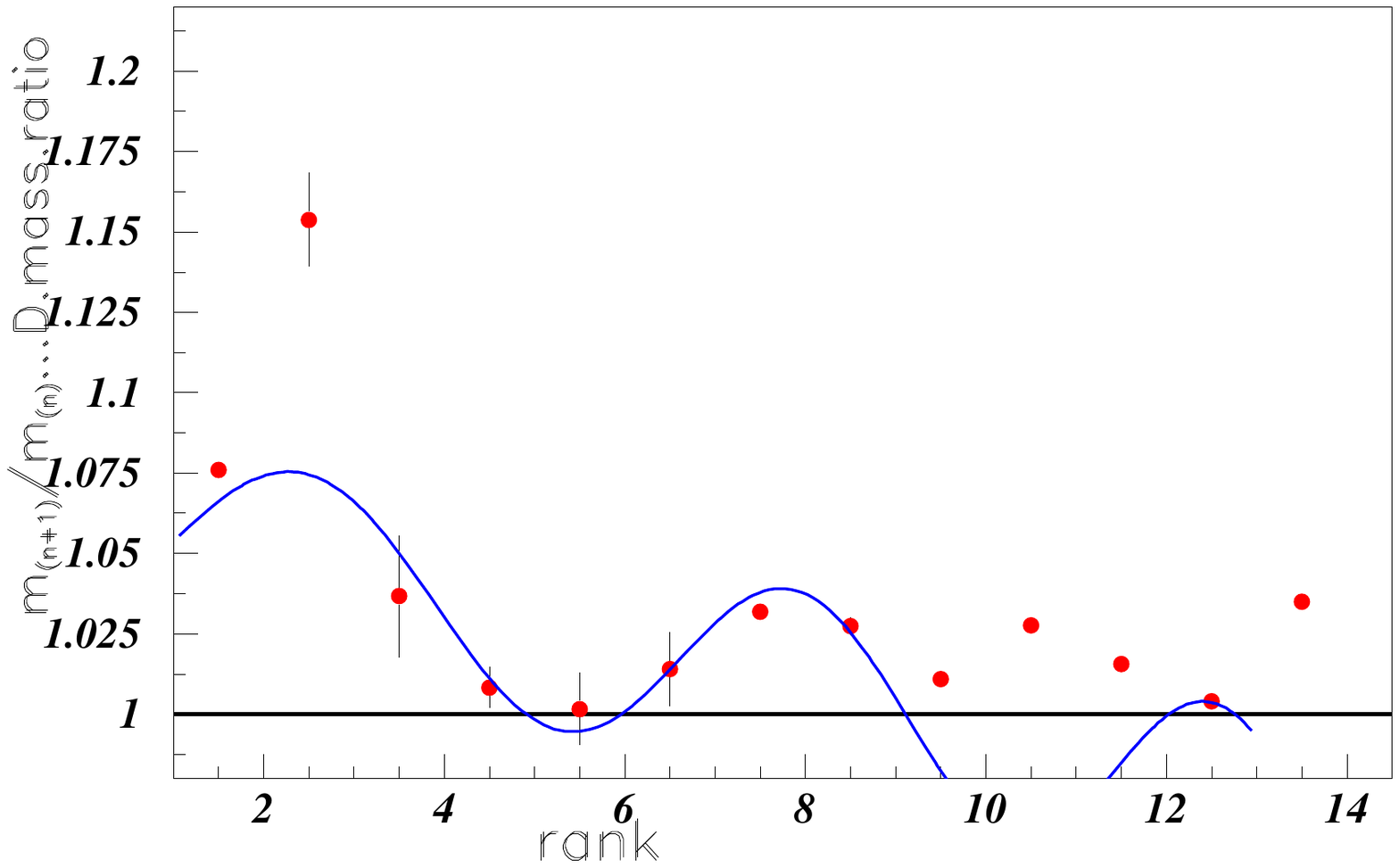}
\end{center}
\end{figure}
\subsection{$s-{\bar s}$ meson masses}
In addition to the PDG masses of the $s-{\bar s}$ mesons, several masses were observed recently.
A resonance at M = 2175~MeV $J^{P}=1^{-}$ was observed by BABAR \cite{aubert} and was tentatively associated with a $s{\bar s}$ or a
$s{\bar s}s{\bar s}$ state. Indeed, it was observed in the  $\Phi$(1020)f$_{0}$(980) invariant mass spectrum.
A consistent mass and width were also observed by the BES Collaboration  \cite{ablikim} in the same final state. The existence of a state around 2.0~GeV was predicted with the quark content: qs${\bar q}{\bar s}$ \cite{hua}. This resonance was sometimes associated with an exotic, hybrid resonance \cite{olsen}.
A narrow resonance in the system $K_{s}K_{s}$ was observed \cite{barkov} at M = 1521.5~MeV. 

There is, up to now, no spectroscopy of $s-{\bar s}$ mesons below M = 2.0~GeV.
\subsection{Charmed strange meson masses}
In addition to the charmed strange meson masses given by PDG \cite{pdg}, the following masses was observed by SELEX \cite{selex}: M = 2632.6 $\pm$ 1.6~MeV, and M = 2856.6 $\pm$ 1.5~MeV and M = 2688 $\pm$ 4~MeV, by BABAR \cite{santoro}.

The log-log distribution of the charmed strange meson masses is plotted in figure~8. 
\begin{figure}[ht]
\begin{center}
\caption{Log-log plot of charmed strange meson masses (MeV).}
\hspace*{-3.mm}
\scalebox{1}[0.7]{
\includegraphics[bb=10 230 530 558,clip,scale=0.5]{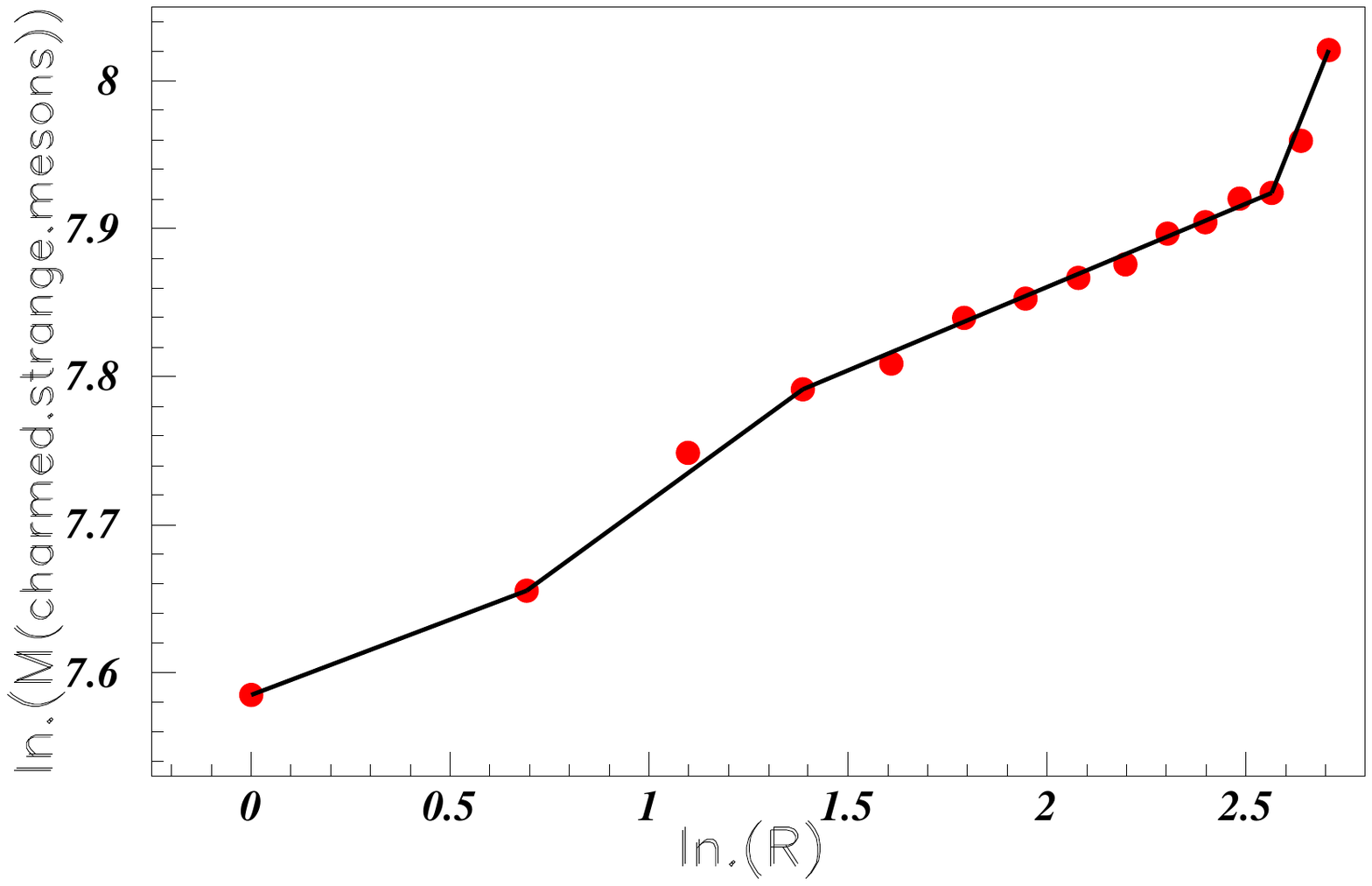}}
\end{center}
\end{figure}
The ratio of $m_{r+1}/m_{r}$ masses for charmed-strange mesons is plotted in figure~9. The peak at low  "r", observed previously in different meson species, is not present here. A  good fit is obtained between the data and calculated distributions. The large value of the last ratio between the 8$^{th}$ mass (M = 2688~MeV) and the 9$^{th}$ mass (M~=~2856~MeV), suggests strongly the missing of (a) charmed strange meson(s), not observed up to now, between these masses.
\begin{figure}[ht]
\begin{center}
\caption{Ratios of $m_{r+1}/m_{r}$ masses, for charmed-strange mesons.}
\hspace*{-3.mm}
\includegraphics[bb=14 230 530 558,clip,scale=0.5]{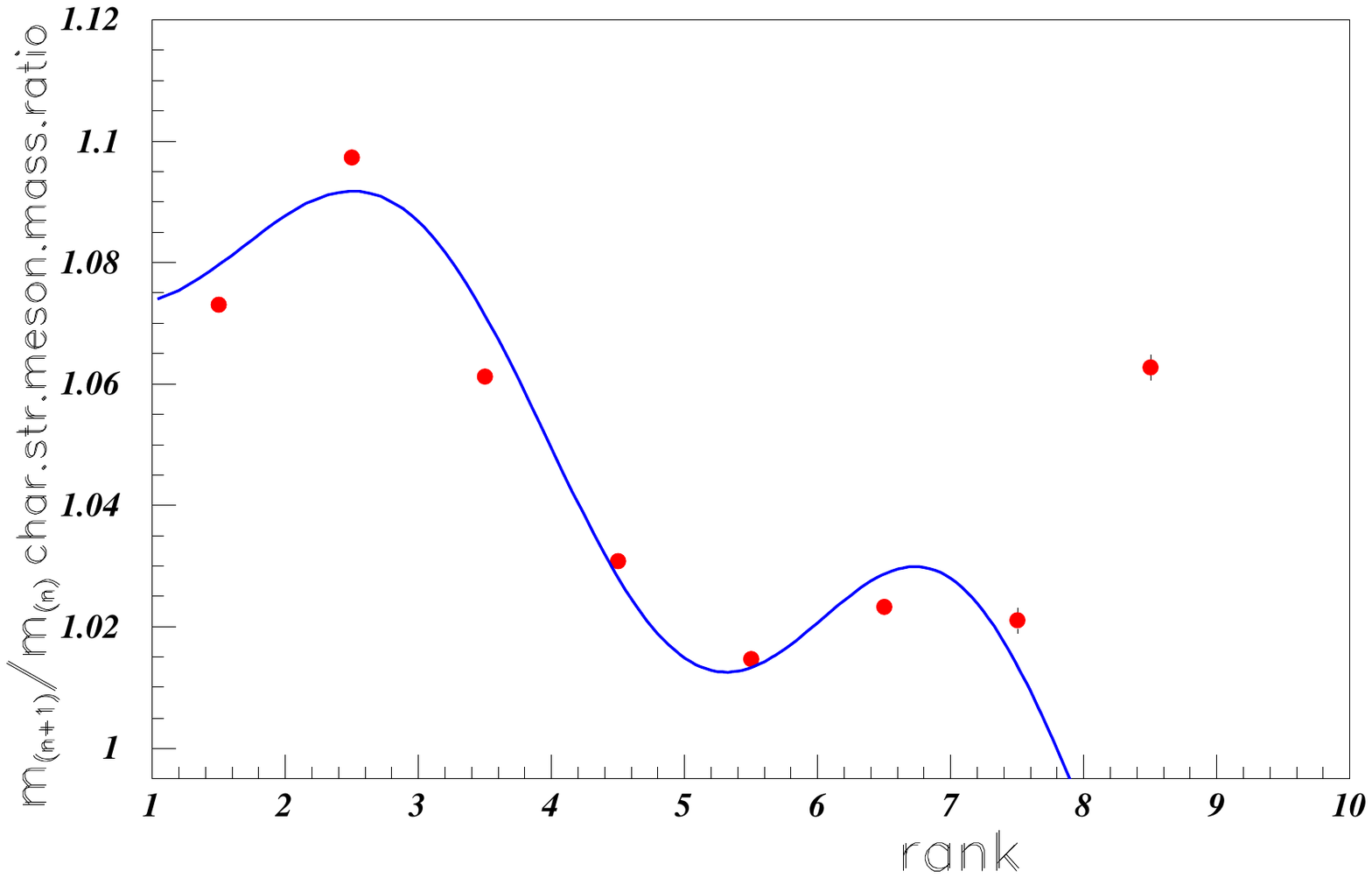}
\end{center}
\end{figure}
\subsection{Bottom meson masses}
The log-log distribution of the bottom meson masses is plotted in figure~10. The small number of bottom mesonic masses, prevents from drawing a mass ratio distribution. From figure~10, we anticipate a single peak centered around r$\approx$~2.5 by comparison with figure 6.
\begin{figure}[ht]
\begin{center}
\caption{Log - log distributions for bottom meson masses (MeV).}
\scalebox{1}[0.7]{
\includegraphics[bb=10 230 530 558,clip,scale=0.5]{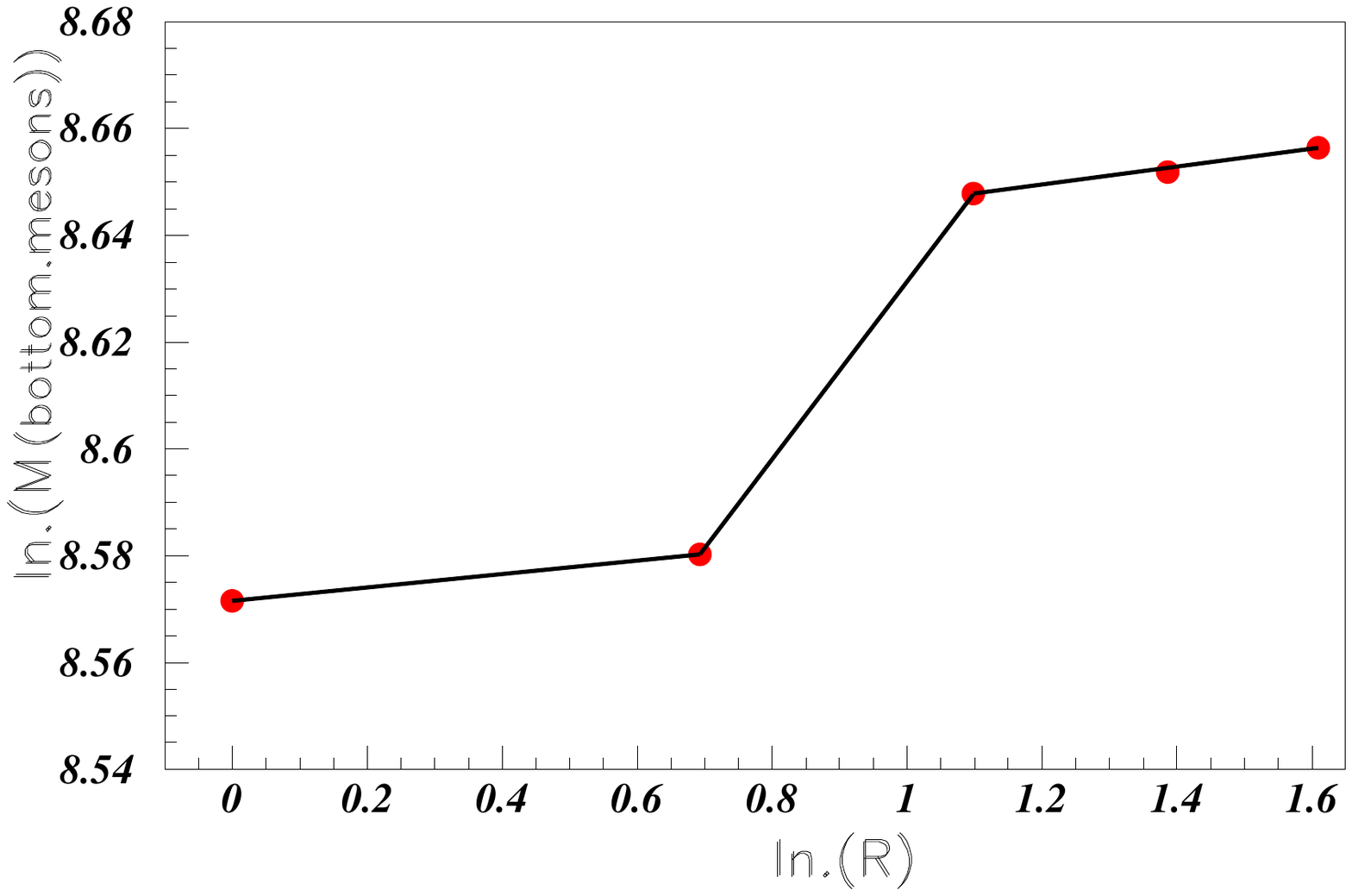}}
\end{center}
\end{figure}
\subsection{Bottom Strange meson masses}
The log-log distribution of the bottom strange meson masses is plotted in figure~11. Here also the same small number of known masses prevent to draw the mass ratio distribution. The masses are rather close to the bottom mesonic masses. A single peak centered around r$\approx$~2.5, is anticipated.
\begin{figure}[ht]
\begin{center}
\caption{Log - log distributions of bottom strange meson masses (MeV)}
\hspace*{-3.mm}
\scalebox{1}[0.7]{
\includegraphics[bb=10 230 530 558,clip,scale=0.5]{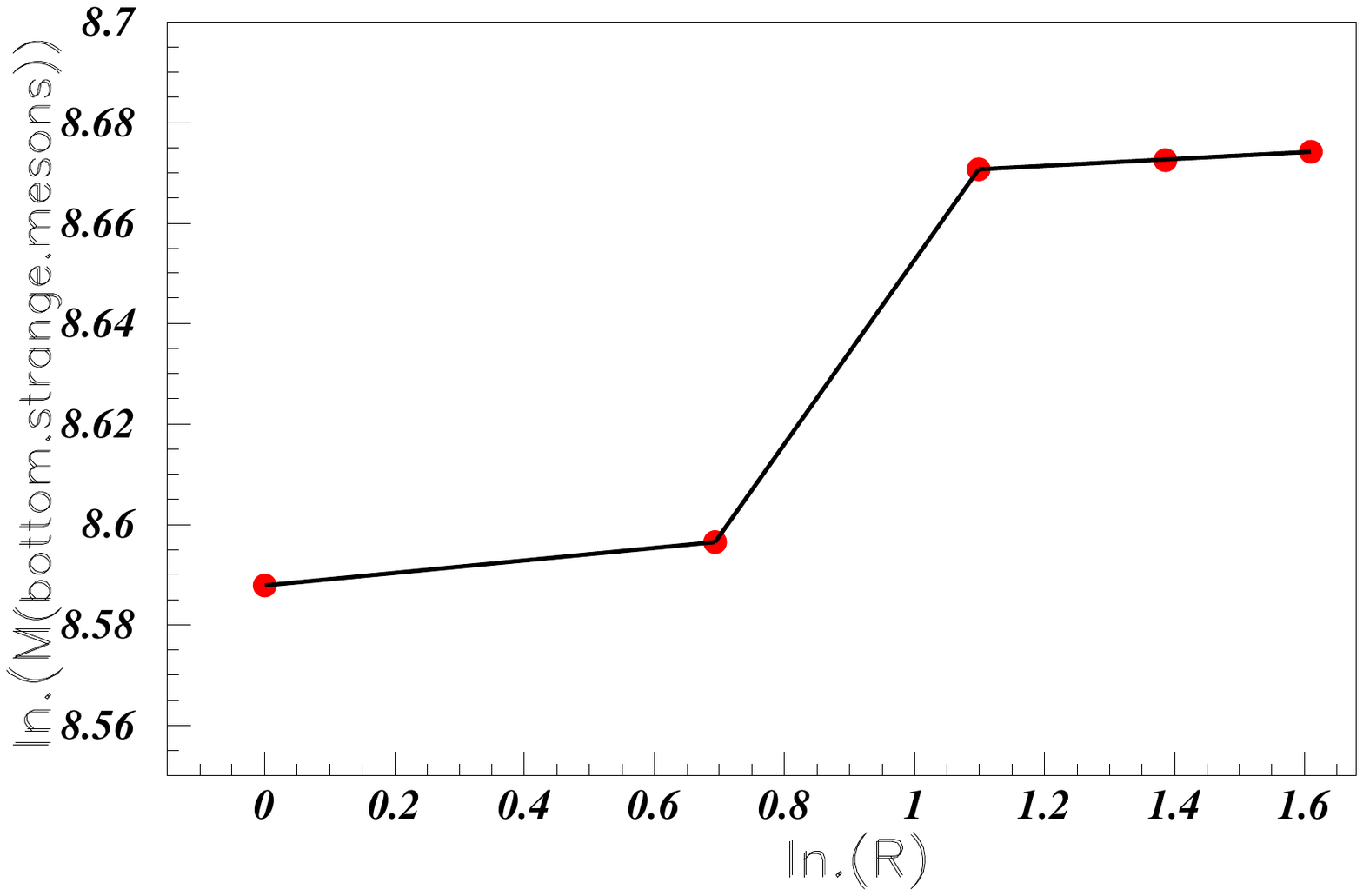}}
\end{center}
\end{figure}

One single bottom charmed meson mass, is reported in \cite{pdg}, namely at M~=~6.276$\pm$0.004~GeV.
\subsection{c - ${\bar c}$ meson masses}
In addition to the masses of $c-{\bar c}$ mesons given by PDG \cite{pdg},
several new masses were observed at M~$\approx$~4.55~GeV \cite{br1}, 4.78~GeV \cite{br2}, 4.87~GeV \cite{br2}, 5.09~GeV \cite{br3}, 
$\approx$ 5.30~GeV \cite{br1}, 5.44~GeV \cite{br3}, 5.66~GeV \cite{br3}, and 5.91~GeV \cite{br3}. Here, a large number (25) of precise masses exist. 
The log-log distribution of the bottom strange meson masses is plotted in figure~12. 
\begin{figure}[ht]
\begin{center}
\caption{Log - log distributions of $c-{\bar c}$ meson masses (MeV).}
\hspace*{-3.mm}
\scalebox{1}[0.7]{
\includegraphics[bb=10 230 530 558,clip,scale=0.5]{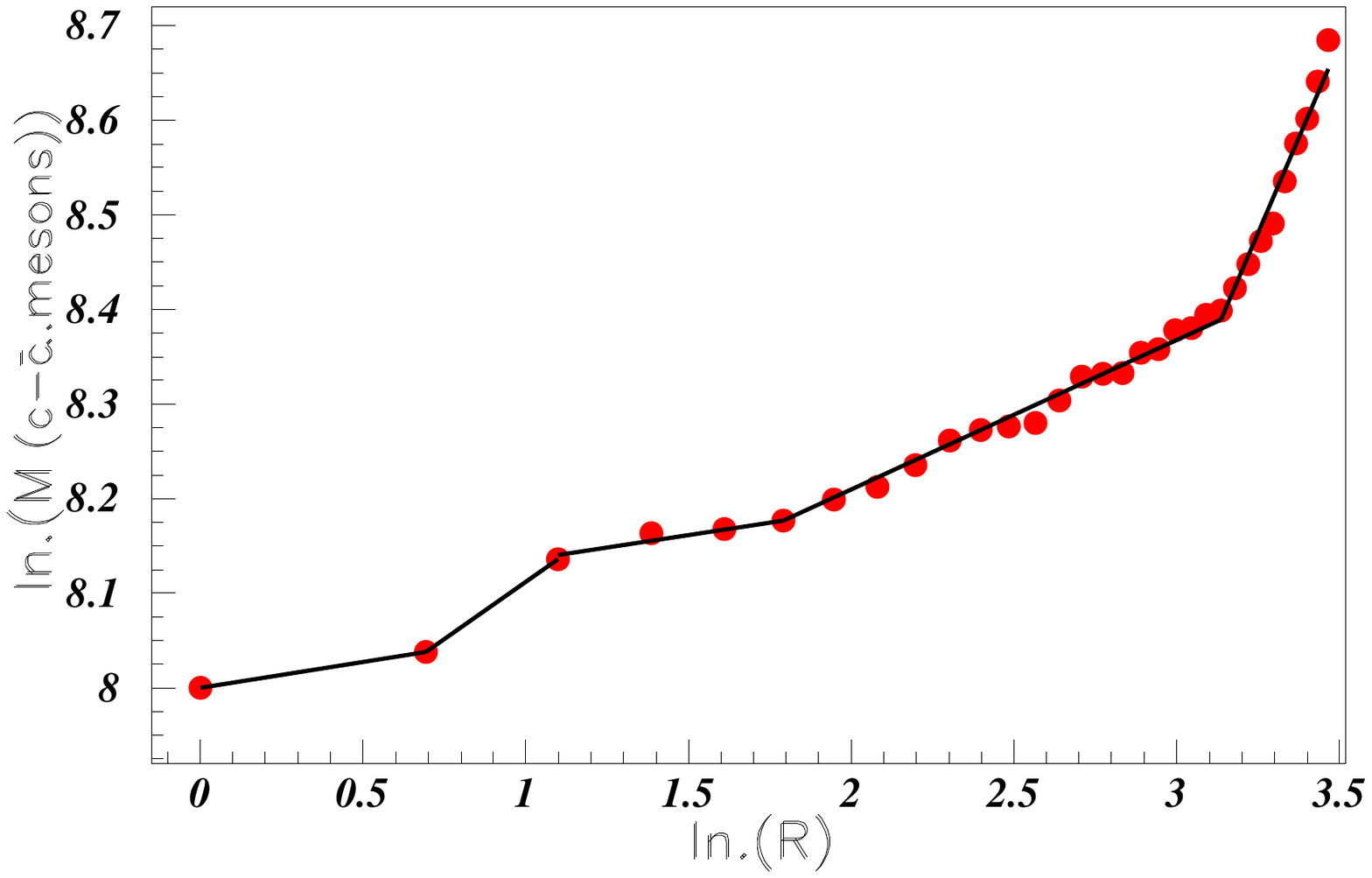}}
\end{center}
\end{figure}
The ratio of $m_{r+1}/m_{r}$ masses, for $c - {\bar c}$ mesons is shown in figure~13. A good fit is obtained, except for the peak at low  "r".
\begin{figure}[ht]
\begin{center}
\caption{Ratios of $m_{r+1}/m_{r}$ masses, for $c - {\bar c}$ mesons.}
\hspace*{-3.mm}
\includegraphics[bb=14 230 530 558,clip,scale=0.5]{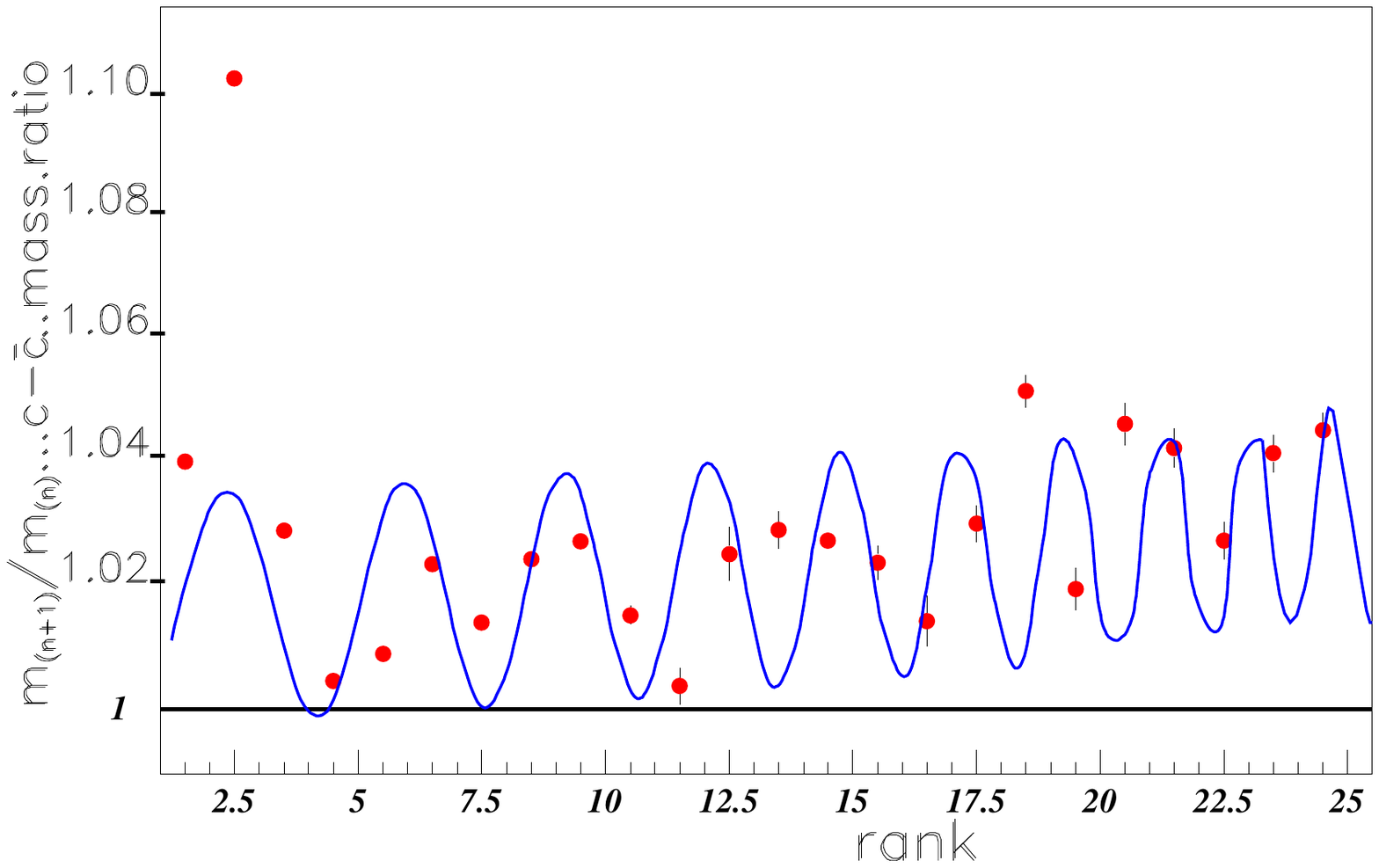}
\end{center}
\end{figure}
The nice agreement  between the data and the fit, obtained for several last ratios, allows us to tentatively predict the next $c - {\bar c}$ meson mass, not observed up to now, to be close to  M~$\approx$~5991~MeV.

\subsection{$b - {\bar b}$  meson masses}
In addition to the masses of $b - {\bar b}$  mesons reported by PDG \cite{pdg}, a resonance at M~=~10.735~GeV was recently reported \cite{br4}. 
The log-log distribution of the bottom strange meson masses is plotted in figure~14. 
\begin{figure}[ht]
\begin{center}
\caption{Log - log distributions of $b-{\bar b}$ meson masses (MeV).}
\hspace*{-3.mm}
\scalebox{1}[0.7]{
\includegraphics[bb=10 230 530 558,clip,scale=0.5]{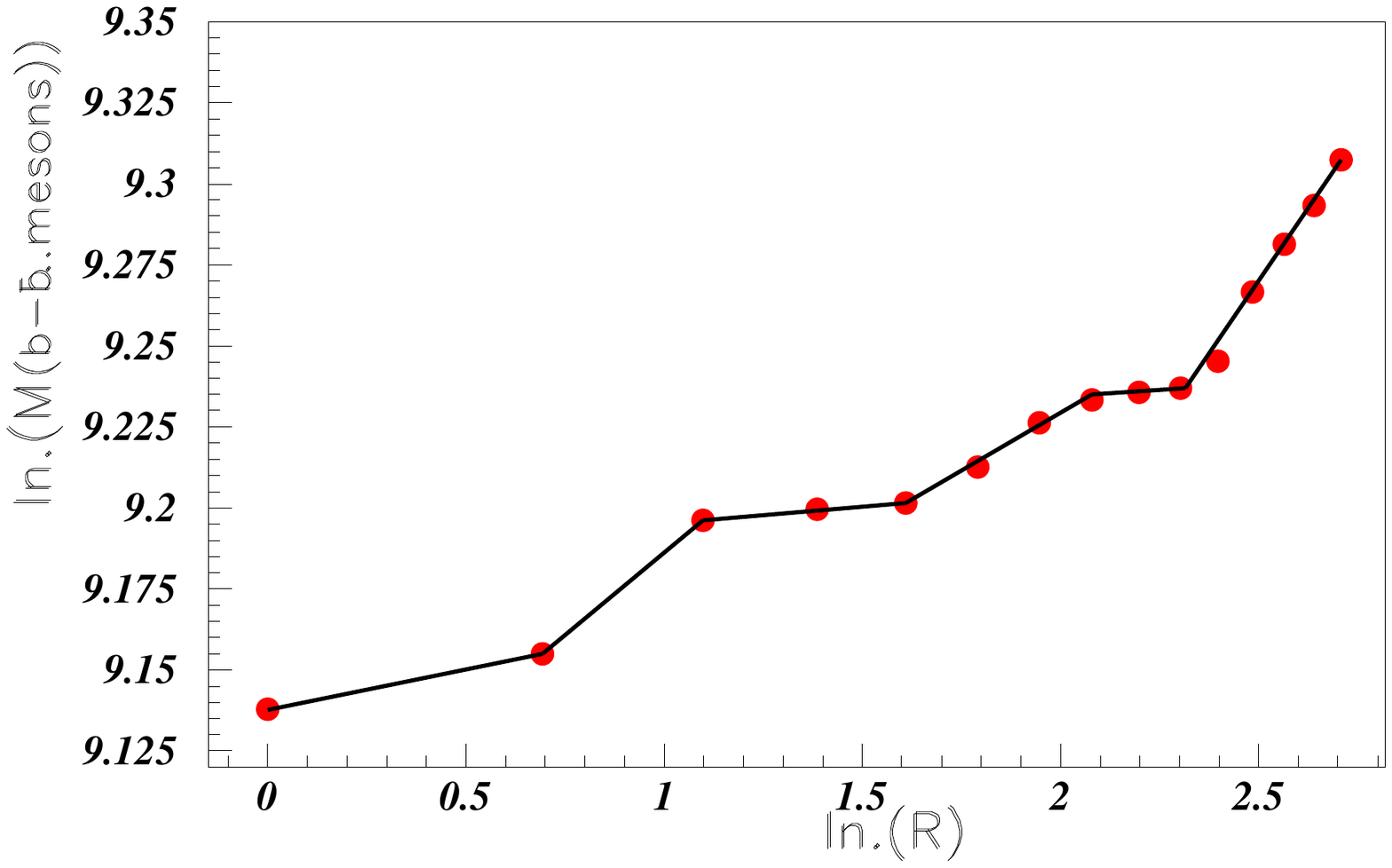}}
\end{center}
\end{figure}
Figure~15 shows the $m_{r+1}/m_{r}$ mass ratios for the
{\it b{$\bar b$}} mesons. We observe a nice fit for all points, except the first ones corresponding to the peak at low  "r". Here again, we tentatively predict the next (16$^{th}$)
 $b-{\bar b}$ meson mass to be close to M~$\approx$~11294~MeV.
\begin{figure}[ht]
\begin{center}
\caption{Ratios of $m_{r+1}/m_{r}$ masses, for $b - {\bar b}$ mesons.}
\hspace*{-3.mm}
\includegraphics[bb=14 230 530 558,clip,scale=0.5]{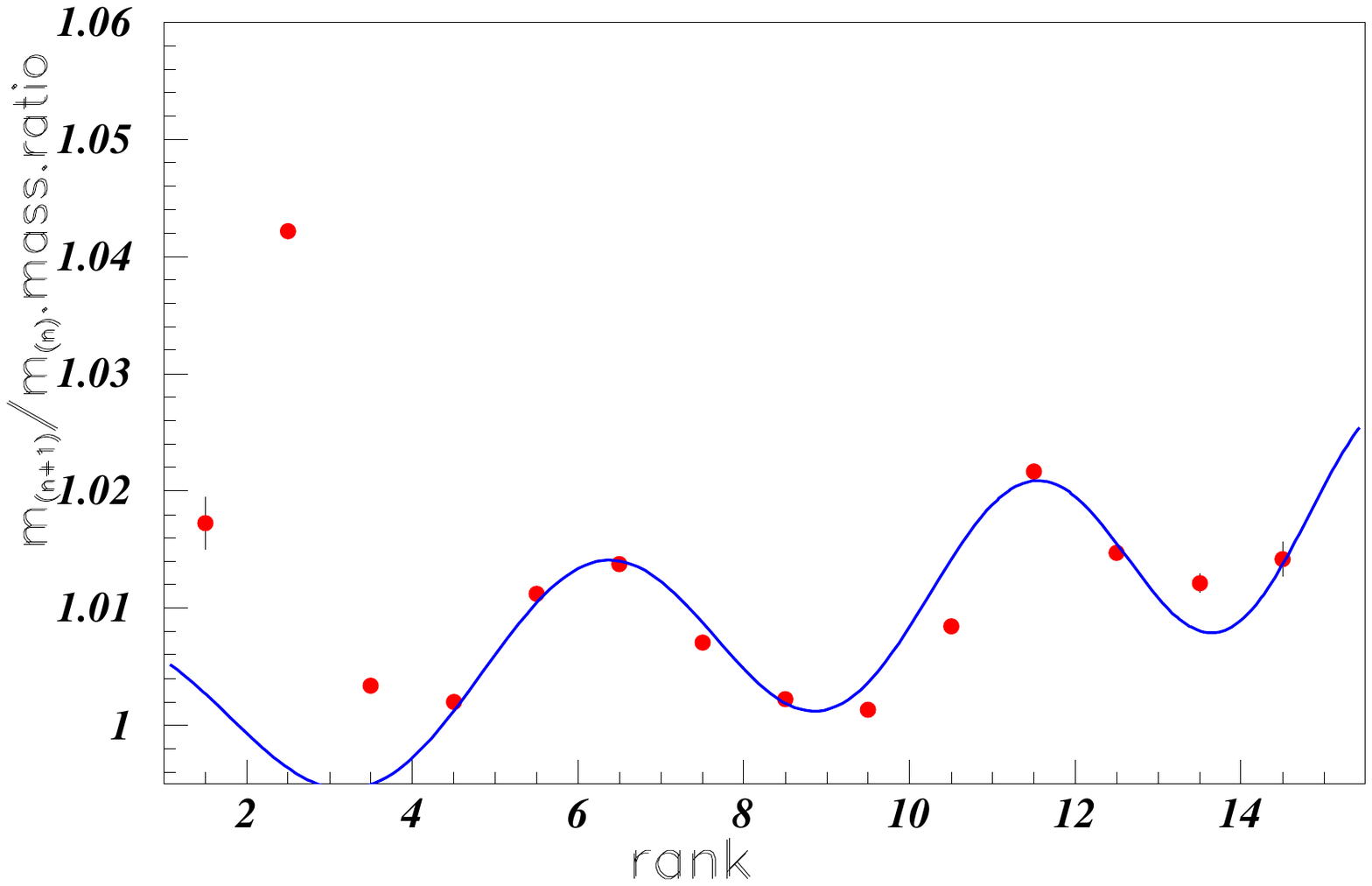}
\end{center}
\end{figure}

Some other heavy flavor mesonic masses exist; they are too scarce to allow a study of their mass variation. A ${\it b{\bar c}}$ meson mass was reported at M$_{B_{c}}$ = 6275.6 $\pm$ 2.9~MeV \cite{giurgiu}.

Many papers discussed the possibility to associate some mesons with hybrids or molecules \cite{lipkin}. Several candidates were presented, all at masses higher than those studied in this paper. For example, the Y(4660) was reported as an 
$f_{0}(980)\Psi'$ bound state \cite{guo}; the decays of $B^{+} \to X(3872)IK^{+}$
and $B^{0} \to X(3872)IK^{0}$ were observed with BABAR at PEP-II \cite{aubert9};
the $\Psi(2S)$ decay to $J/\Psi$ was studied at BES \cite{ablikim9}; the $Z^\pm(4430)$ was observed at BELLE \cite{abey} and discussed as an excellent candidate for being an exotic state \cite{ke}.
\section{PDG baryon Masses}
\subsection{N Baryon masses}
The log-log distribution of the N baryon masses is plotted in figure~16.
\begin{figure}[ht]
\begin{center}
\caption{Log - log distributions for the N baryonic masses (MeV).}
\hspace*{-3.mm}
\scalebox{1}[0.7]{
\includegraphics[bb=10 230 530 558,clip,scale=0.5]{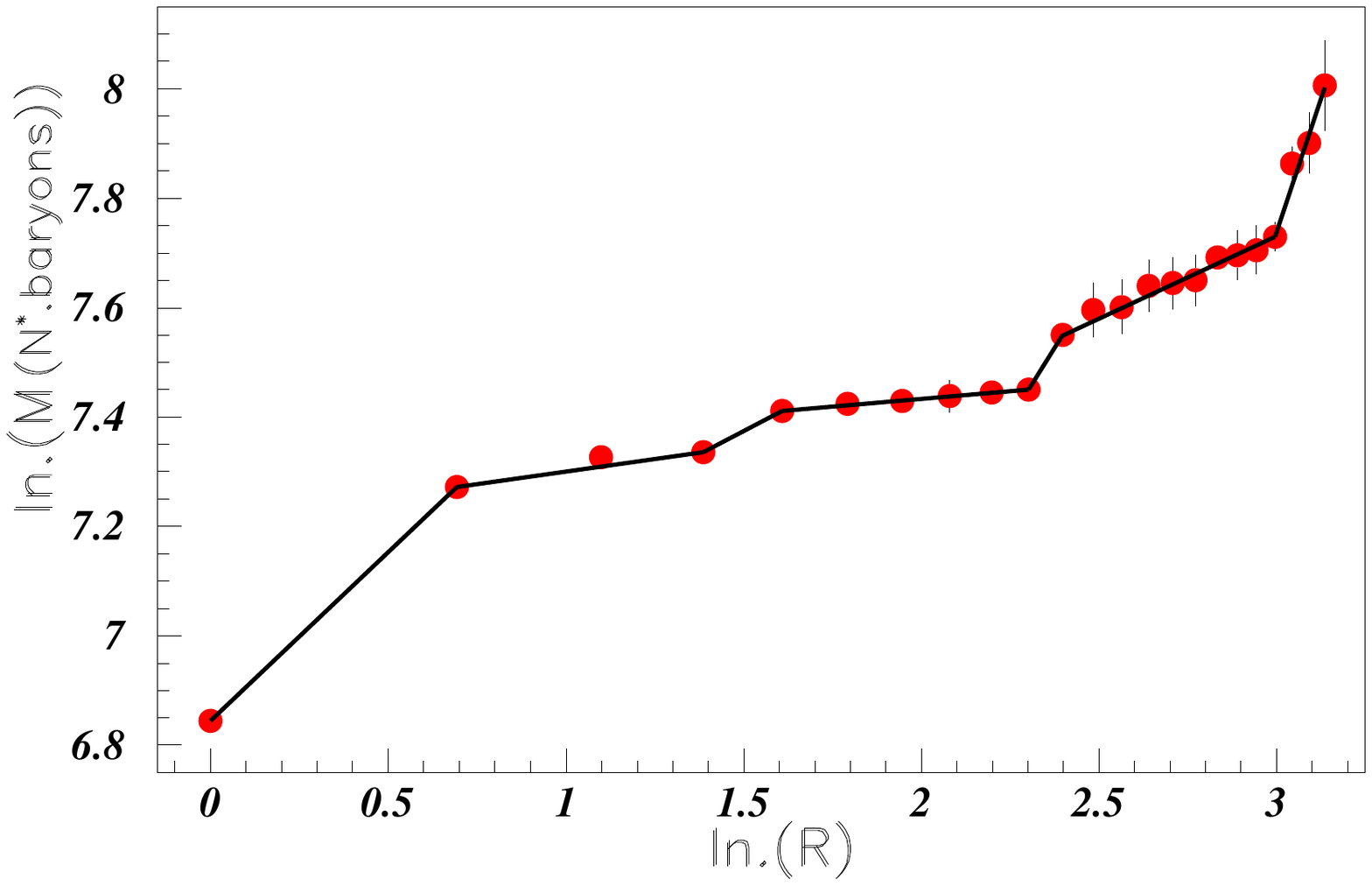}}
\end{center}
\end{figure}
The figure shows several straight lines, with a possible gap  between the 10$^{th}$ and 11$^{th}$ masses (M = 1720~MeV and 2190~MeV). These gaps manifest themselves by a little jump with several points aligned after the jump. The ratio of $m_{r+1}/m_{r}$ masses, for N baryons is shown in figure~17.
\begin{figure}[ht]
\begin{center}
\caption{Ratios of $m_{r+1}/m_{r}$ masses, for N baryonic masses.}
\hspace*{-3.mm}
\includegraphics[bb=14 230 530 558,clip,scale=0.5]{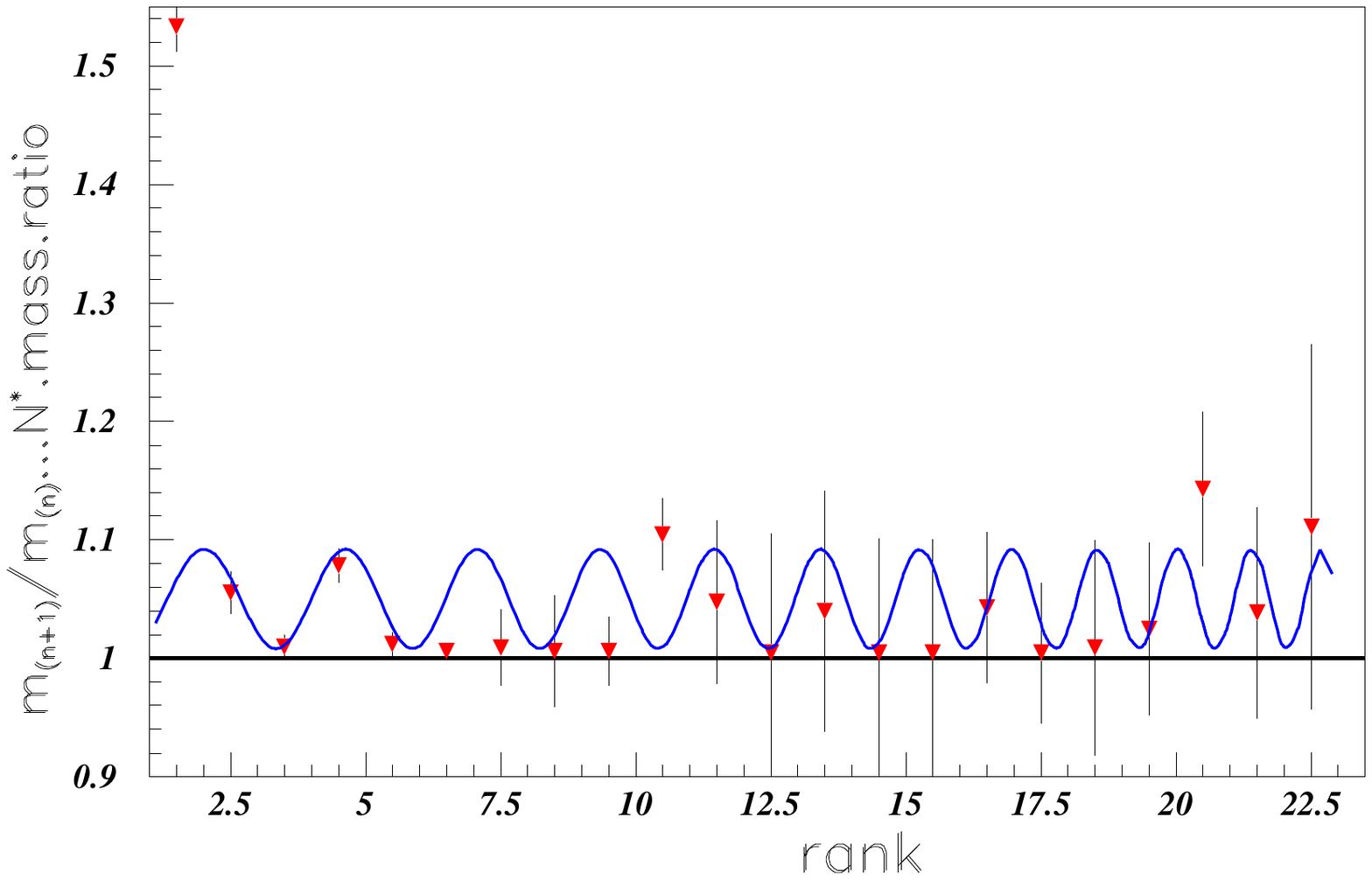}
\end{center}
\end{figure}
The unprecision on the unflavoured baryonic masses are large, starting at M~=~1700~MeV. Here too, the large  first value is not correctly fitted.
\subsection{$\Delta$ baryon masses}
The log - log distribution of the $\Delta$ baryon masses is plotted in figure~18. 
\begin{figure}[ht]
\begin{center}
\caption{Log - log distributions for the $\Delta$ baryonic masses (MeV).}
\hspace*{-3.mm}
\scalebox{1}[0.7]{
\includegraphics[bb=10 230 530 558,clip,scale=0.5]{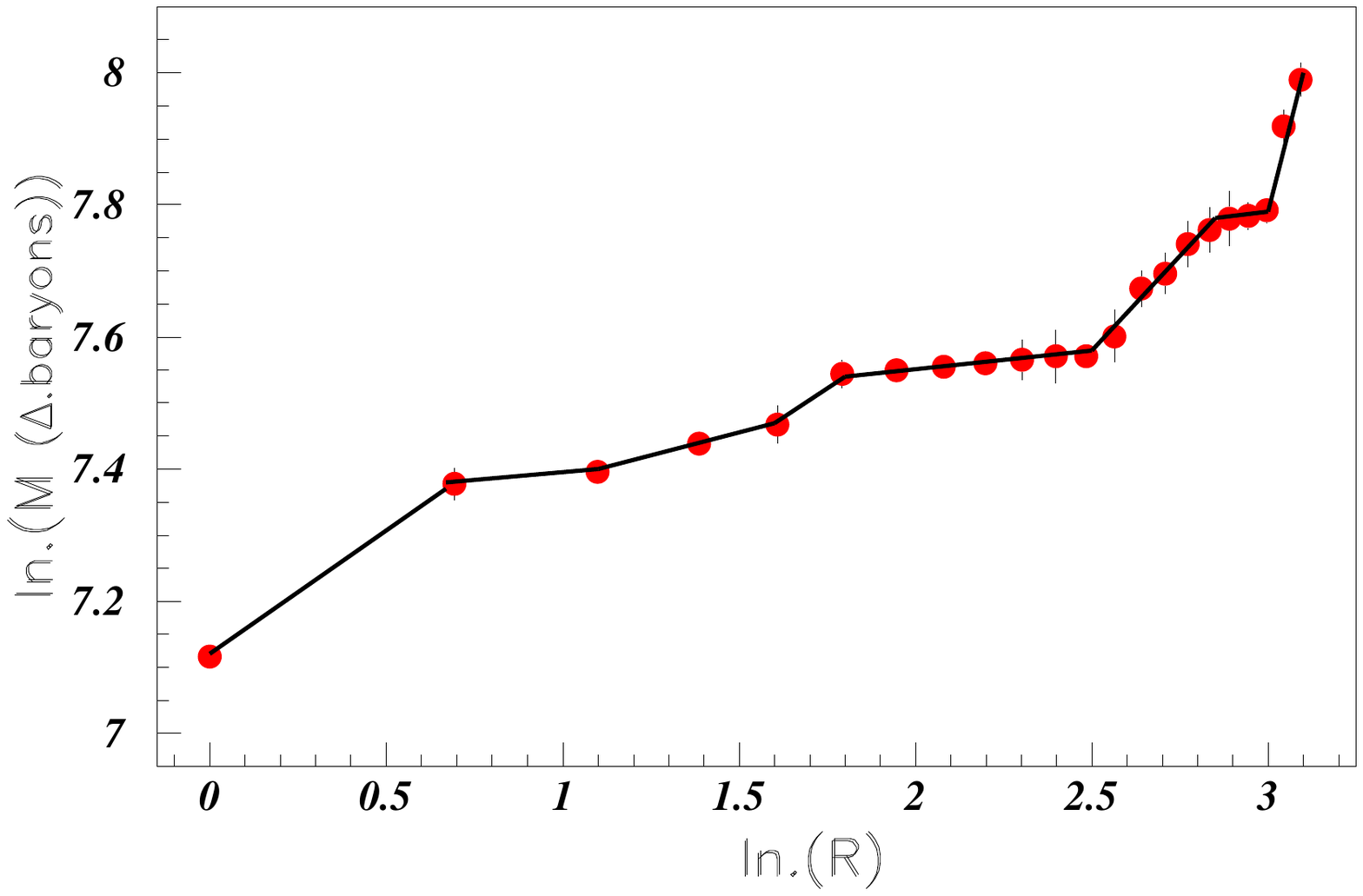}}
\end{center}
\end{figure}
From the figure, we can suggest that the mass of the third $\Delta$ baryon,
 M~=~1630~MeV (called $\Delta$(1620)), could be increased by $\Delta$M~$\approx$~20~MeV.  Such small increase is possible, since the masses of the $\Delta $ baryons are rather unprecised. 
The large mass unprecisions, and the proximity of many masses (7 masses between M = 1890~MeV and 1940~MeV), justify the omission of the  $m_{r+1}/m_{r}$ distribution for the $\Delta $ baryons.
\subsection{$\Lambda$ baryon masses}
The log-log distribution of the $\Lambda$ baryon masses is plotted in figure~19. 
\begin{figure}[ht]
\begin{center}
\caption{Log - log distributions for the $\Lambda$ baryonic masses (MeV).}
\hspace*{-3.mm}
\scalebox{1}[0.7]{
\includegraphics[bb=10 230 530 558,clip,scale=0.5]{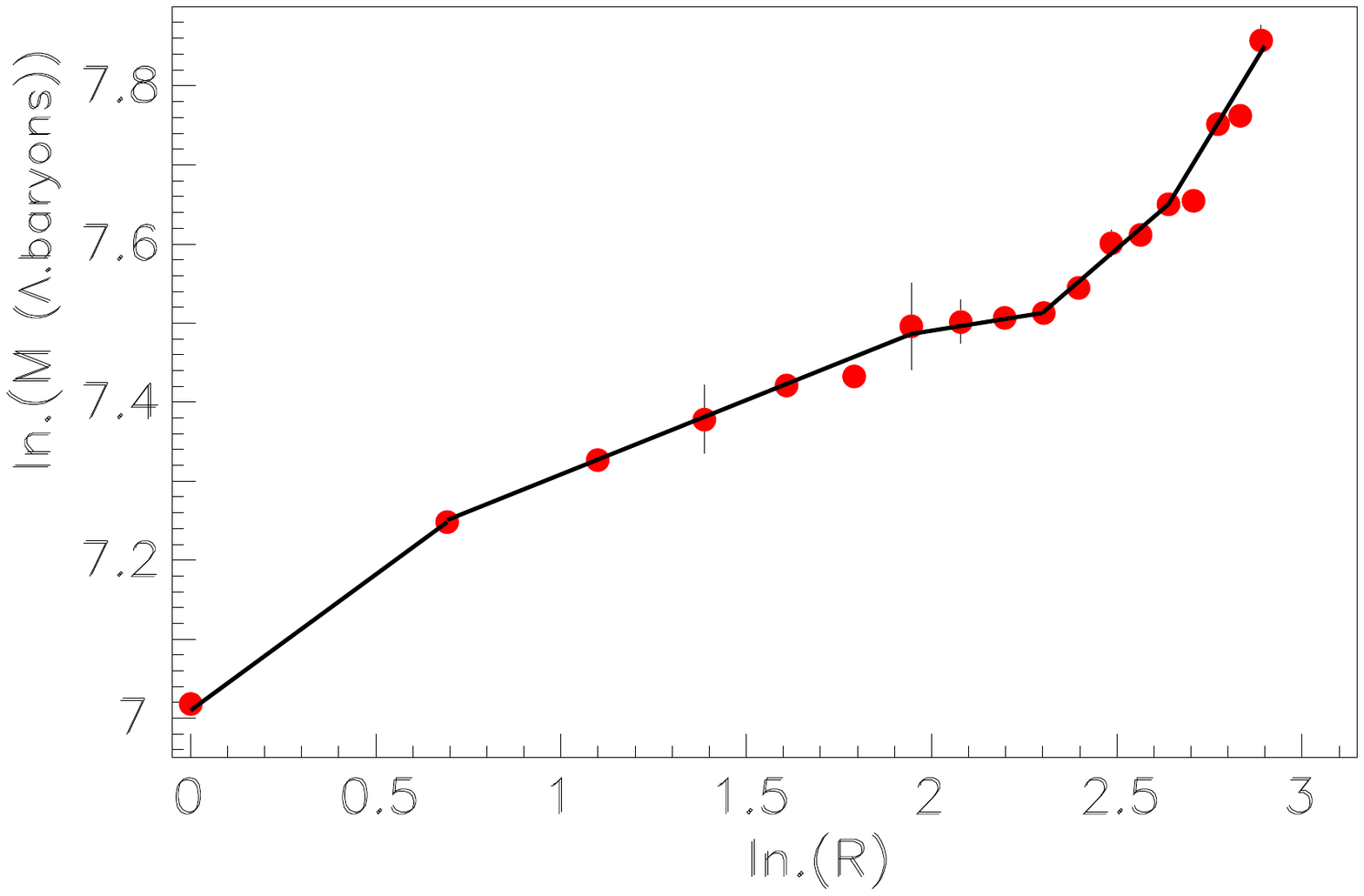}}
\end{center}
\end{figure}
There is a gap between the 6$^{th}$ and the 7$^{th}$ masses 
(M~=~1690~MeV and 1800~MeV), but it is not clear that it corresponds to a missing  $\Lambda$ baryon mass. M~=~1690~MeV could be a little low (see the 6$^{th}$ and the 7$^{th}$ points in figure 19). The ratio of $m_{r+1}/m_{r}$ masses, for $\Lambda$ baryons is shown in figure~20.
\begin{figure}[ht]
\begin{center}
\caption{Ratios of $m_{r+1}/m_{r}$ masses, for $\Lambda$ baryons.}
\hspace*{-3.mm}
\includegraphics[bb=14 230 530 558,clip,scale=0.5]{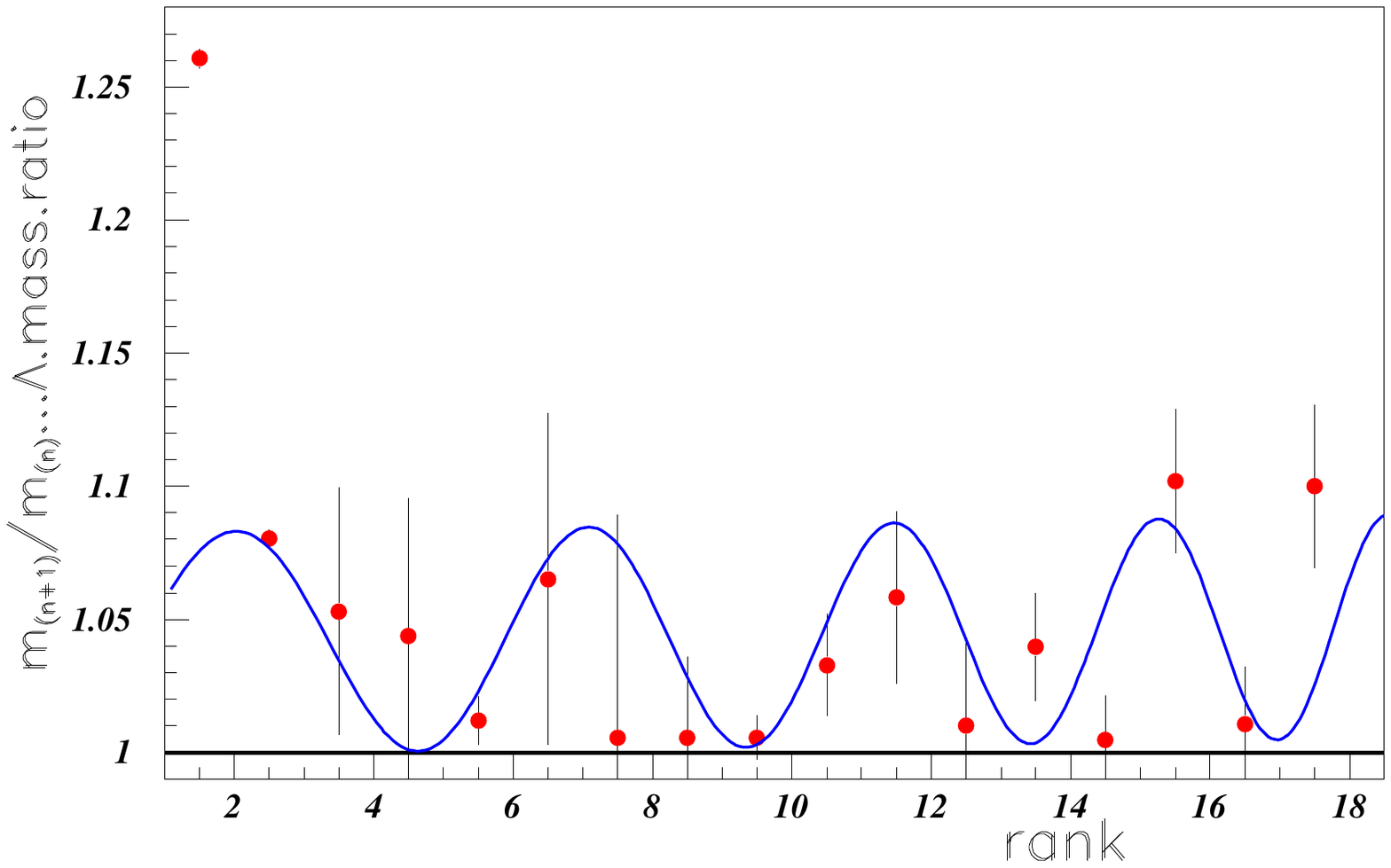}
\end{center}
\end{figure}
Except the first large data, outside the fit, the agreement between the fitted and the experimental distributions is obtained, but is of poor value due to large error bars.
\subsection{$\Sigma$ baryon masses}
\begin{figure}[ht]
\begin{center}
\caption{Log - log distributions for the $\Sigma$ baryonic masses (MeV).}
\hspace*{-3.mm}
\scalebox{1}[0.7]{
\includegraphics[bb=10 230 530 558,clip,scale=0.5]{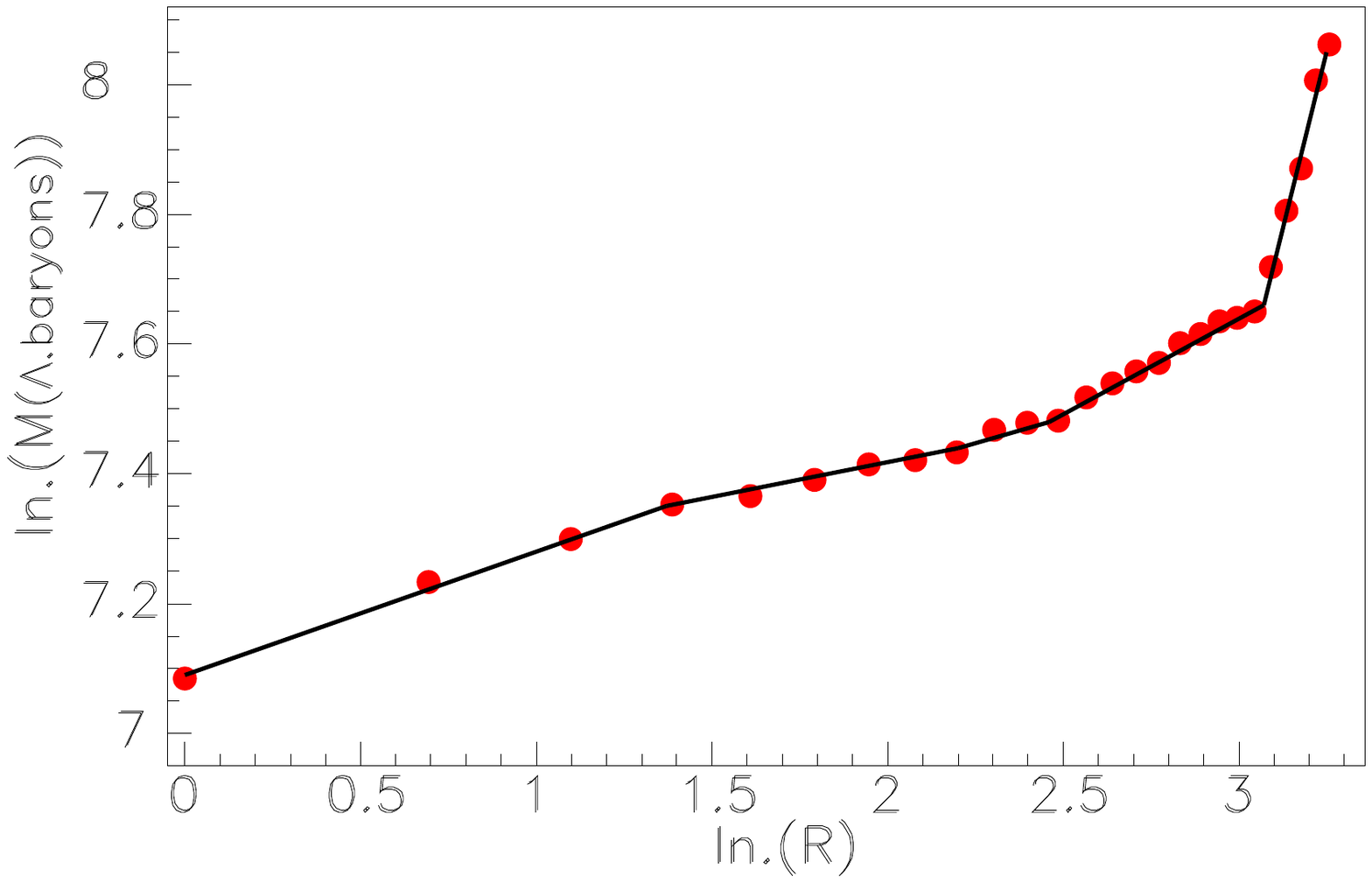}}
\end{center}
\end{figure}
The log-log distribution of the $\Sigma$ baryon masses is plotted in  figure~21. 
 The ratio of $m_{r+1}/m_{r}$ masses, for $\Sigma$ baryons is shown in figure~22. 
\begin{figure}[ht]
\begin{center}
\caption{Ratios of $m_{r+1}/m_{r}$ masses, for $\Sigma$ baryon masses.}
\hspace*{-3.mm}
\includegraphics[bb=14 230 530 558,clip,scale=0.5]{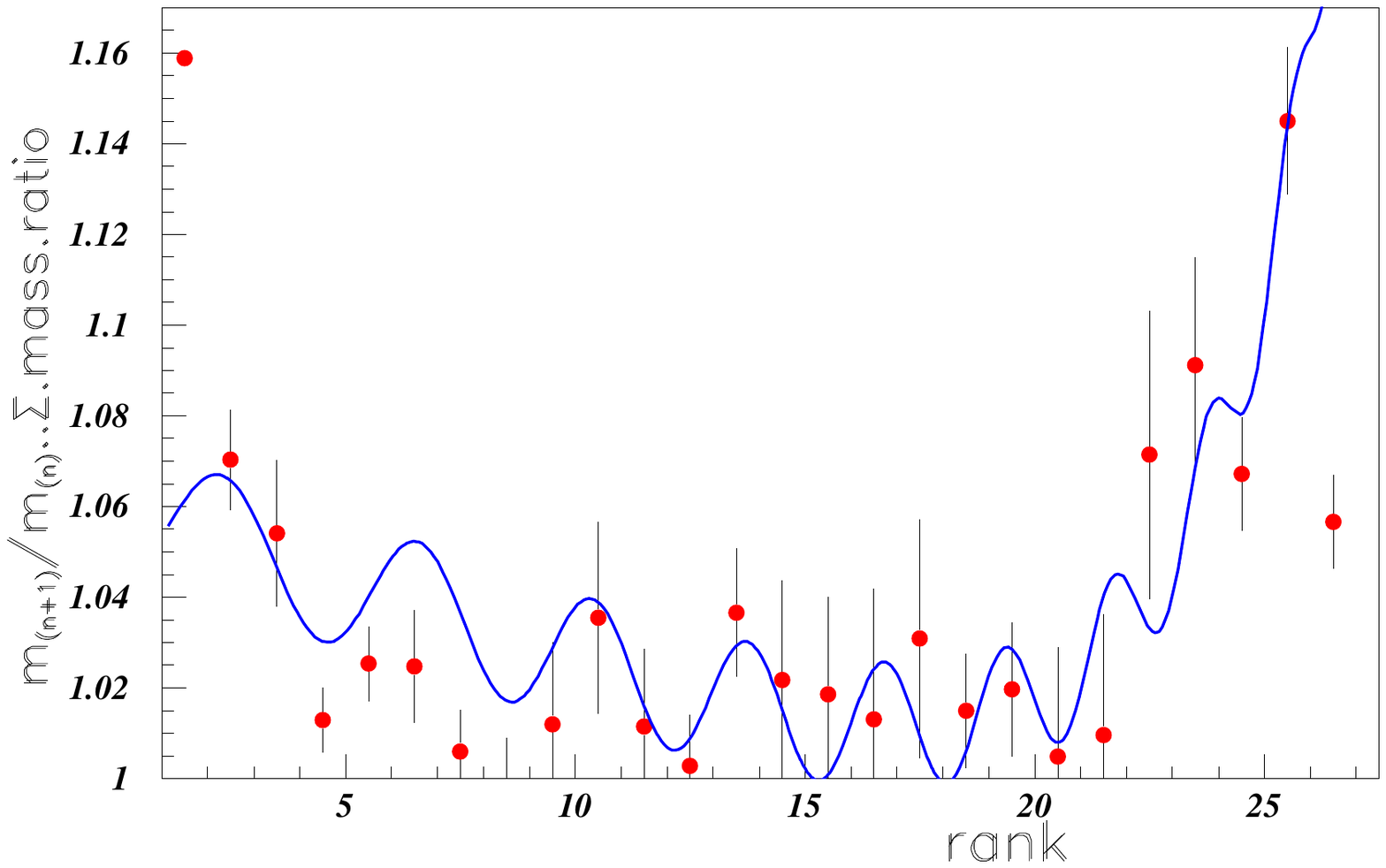}
\end{center}
\end{figure}
The first large peak at low "r" is not reproduced by the fit; moreover the large error bars make the agreement between the fitted and the experimental distributions valueless. The increase of data points for rank larger than 21, may be due to missing $\Sigma$ baryons, still not observed. The fit for the actual data, shown in figure~22, is obtained with a doubly equation (1) with different parameters. These five last data points correspond to the last points in  figure~21 showing the log-log distribution of $\Sigma$ baryons. The alignement of these data in figure 21, showing a large slope, is totally different from the alignement of the  data for smaller rank. These last data show clearly the need for discrete scale invariance, with different parameters describing the different range of the distribution. 
\subsection{$\Xi$ baryon masses}
\begin{figure}[ht]
\begin{center}
\caption{Log - log distributions for the $\Xi$ baryonic masses (MeV).}
\hspace*{-3.mm}
\scalebox{1}[0.7]{
\includegraphics[bb=10 230 530 558,clip,scale=0.5]{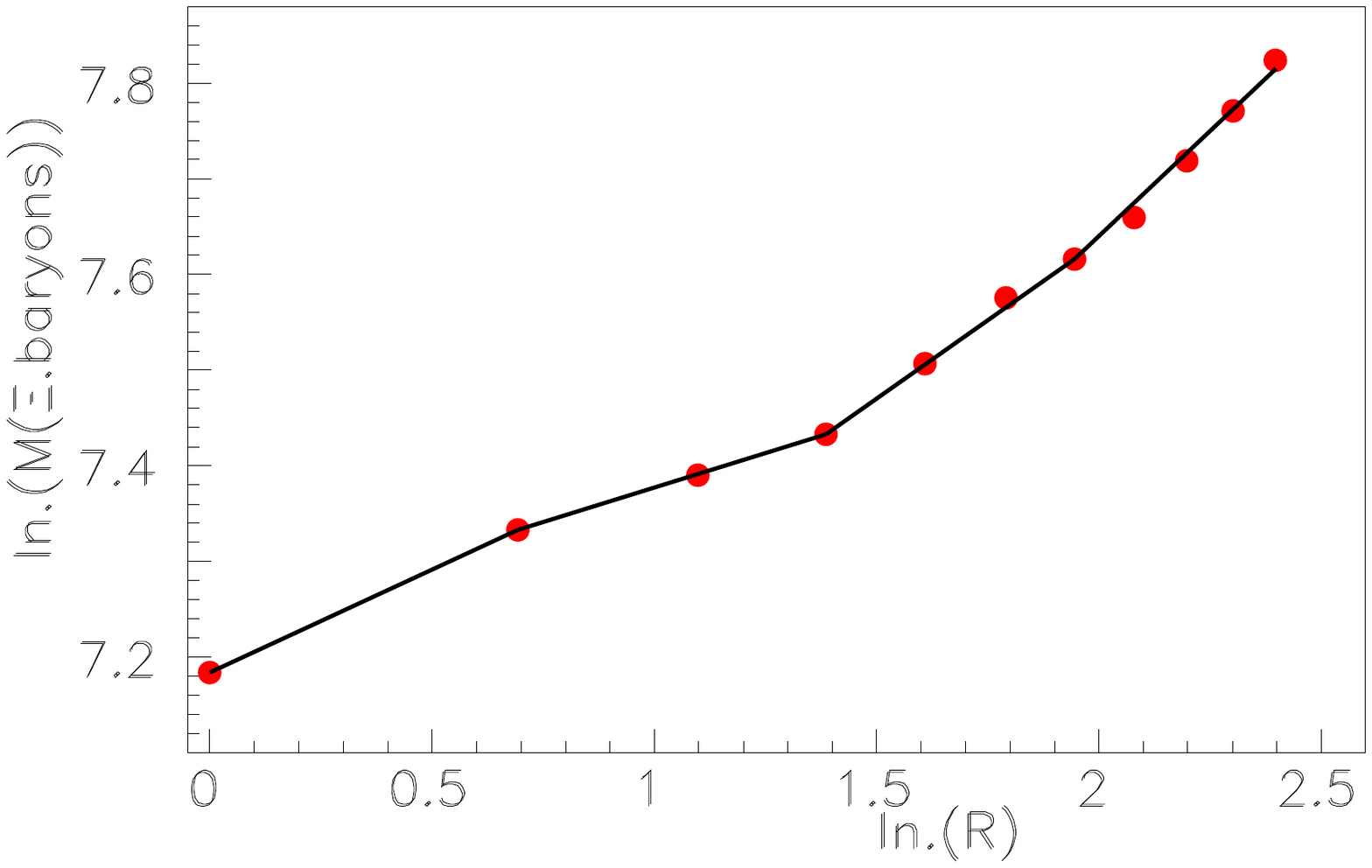}}
\end{center}
\end{figure}
The log-log distribution of the $\Xi$ baryon masses is plotted in 
figure~23. 
 The ratio of $m_{r+1}/m_{r}$ masses, for $\Xi$ baryons is shown in figure~24. 
\begin{figure}[ht]
\begin{center}
\caption{Ratios of $m_{r+1}/m_{r}$ masses, for  $\Xi$ baryonic masses.}
\hspace*{-3.mm}
\includegraphics[bb=14 230 530 558,clip,scale=0.5]{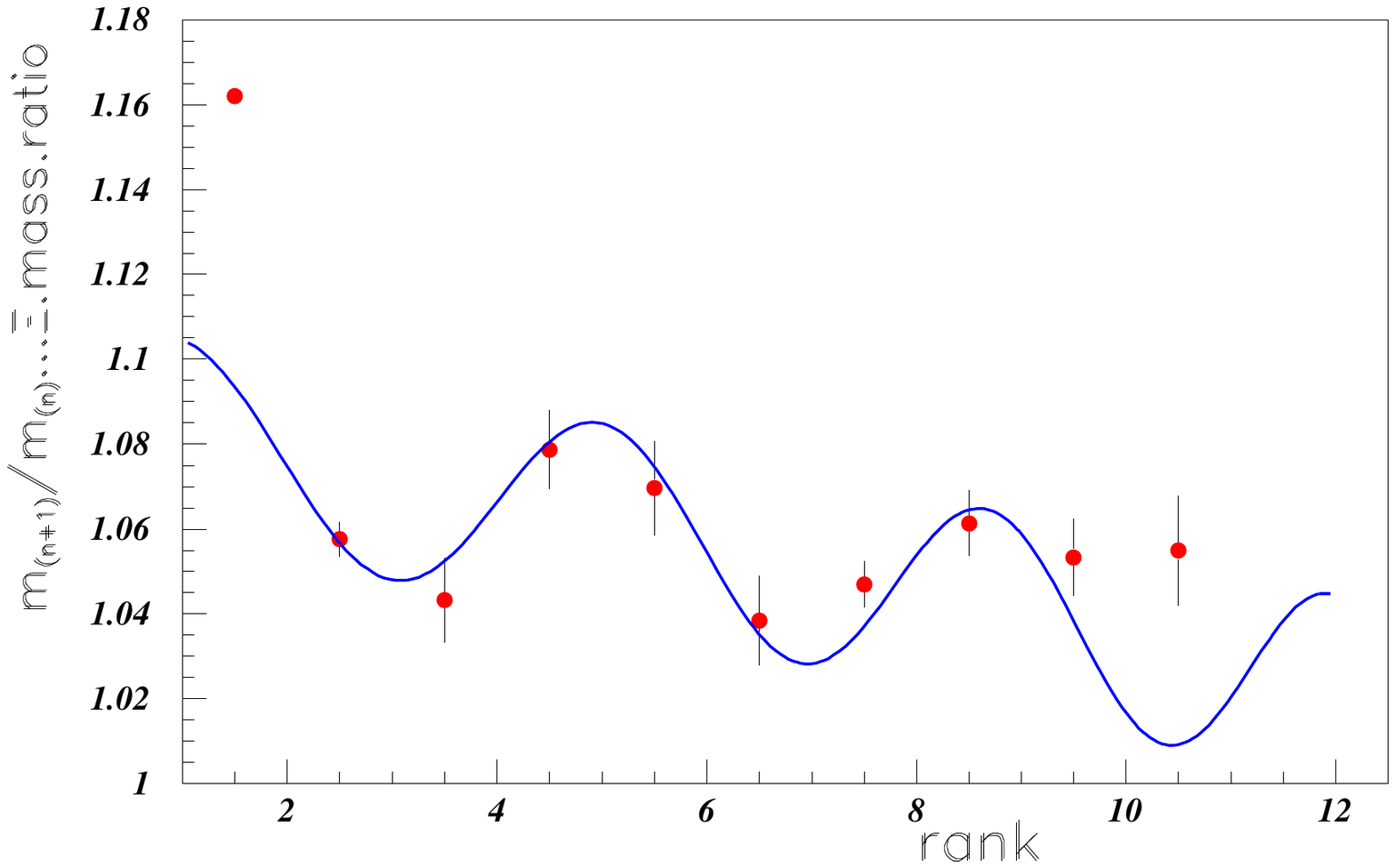}
\end{center}
\end{figure}
The error bars are smaller than previously observed in baryonic species. We also observe a good fit for the complete distribution, except the first point, as usually, and except the last point allowing us to anticipate the possible absence of (a) $\Xi$ baryonic mass(es) between M = 2370~MeV and 2500~MeV.
\subsection{$\Omega$ baryon masses}
Only four $\Omega$ baryonic masses are reported in PDG \cite{pdg} at 
M~=~1672.45~MeV, 2252~MeV, 2380~MeV, and 2470~MeV, the last two being "omitted from summary table". They are too scarce to be considered in the frame of our study.
\subsection{Charmed  baryon masses}
\begin{figure}[ht]
\begin{center}
\caption{Log - log distributions for the  charmed  baryonic masses (MeV).}
\hspace*{-3.mm}
\scalebox{1}[0.7]{
\includegraphics[bb=10 230 530 558,clip,scale=0.5]{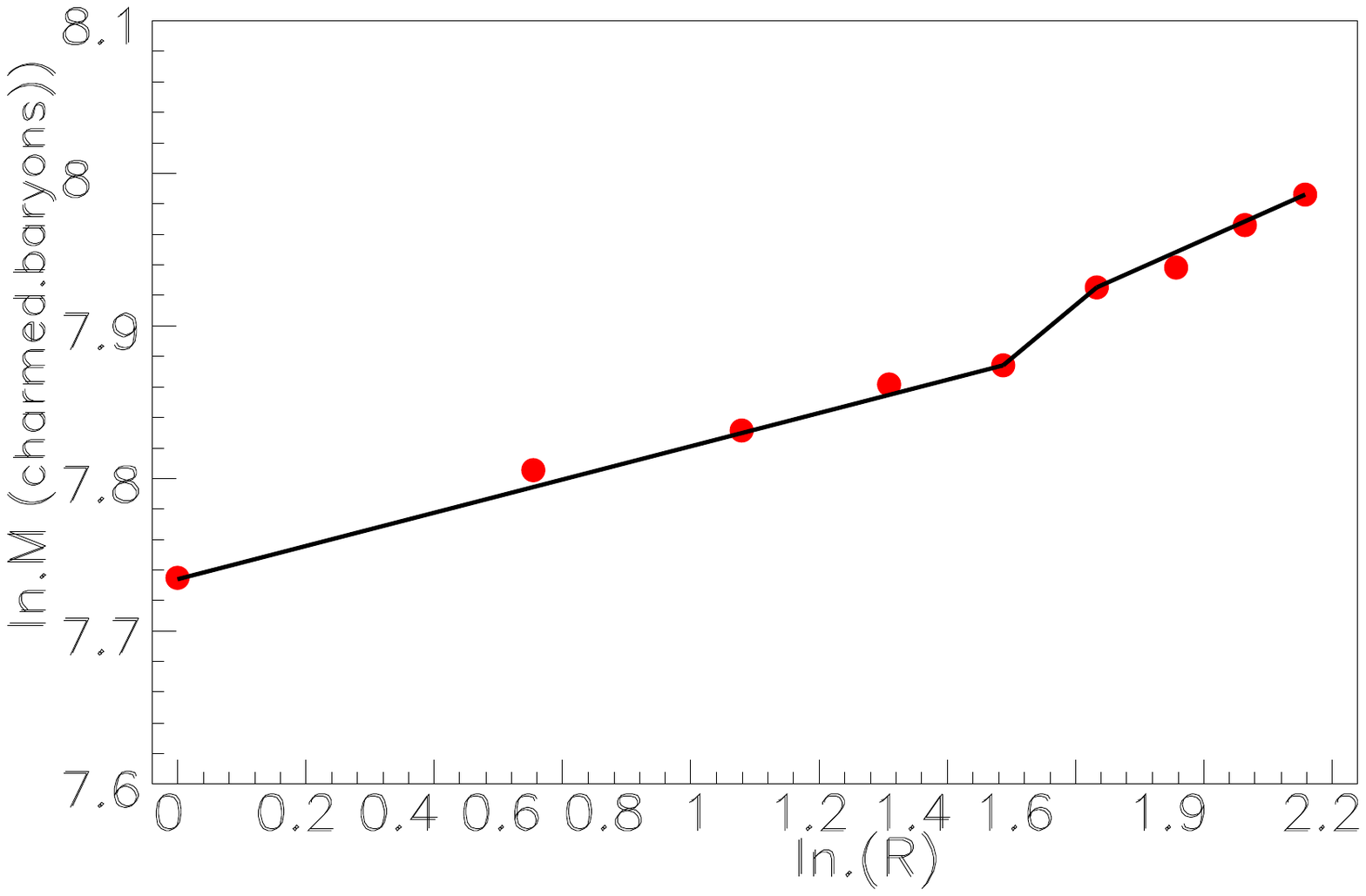}}
\end{center}
\end{figure}
Only 6 masses of charmed $\Lambda$  baryons are reported in PDG \cite{pdg} and 3 masses of charmed $\Sigma$ baryons. Therefore, we analyzed simultaneously the charmed $\Lambda$  baryons and the charmed $\Sigma$ baryons.
The log-log distribution of the charmed  baryonic masses is plotted in figure~25. 
\begin{figure}[ht]
\begin{center}
\caption{Ratios of $m_{r+1}/m_{r}$ masses, for charmed baryonic masses.}
\hspace*{-3.mm}
\includegraphics[bb=14 230 530 558,clip,scale=0.5]{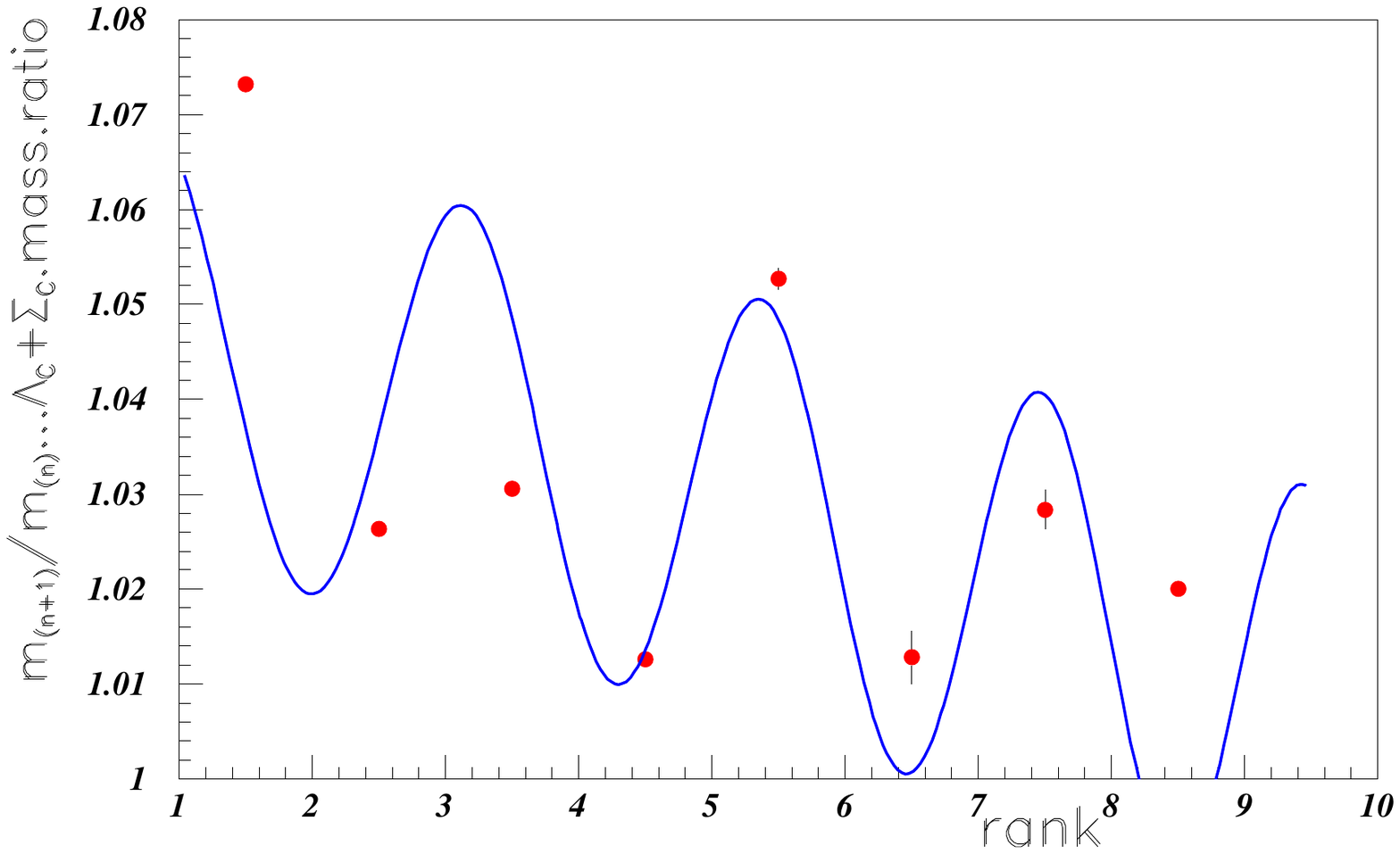}
\end{center}
\end{figure}
 The ratio of $m_{r+1}/m_{r}$ masses, for the charmed  baryons is shown in figure~26. The error bars are small. 
We also observe a rather good fit for the complete distribution, except the first point, as usually, and except for the last point allowing us to anticipate the possible absence of (a) charmed baryonic mass(es) between M = 2881.53~MeV and 2939.3~MeV.
\subsection{The $\Xi$ charmed baryon masses}
Ten $\Xi_{c}$  masses are reported in PDG \cite{pdg}.

The log-log distribution of the $\Xi_{c}$ masses is plotted in figure~27.
\begin{figure}[ht]
\begin{center}
\caption{Log - log distributions for the  $\Xi_{c}$  masses (MeV).}
\hspace*{-3.mm}
\scalebox{1}[0.7]{
\includegraphics[bb=10 230 530 558,clip,scale=0.5]{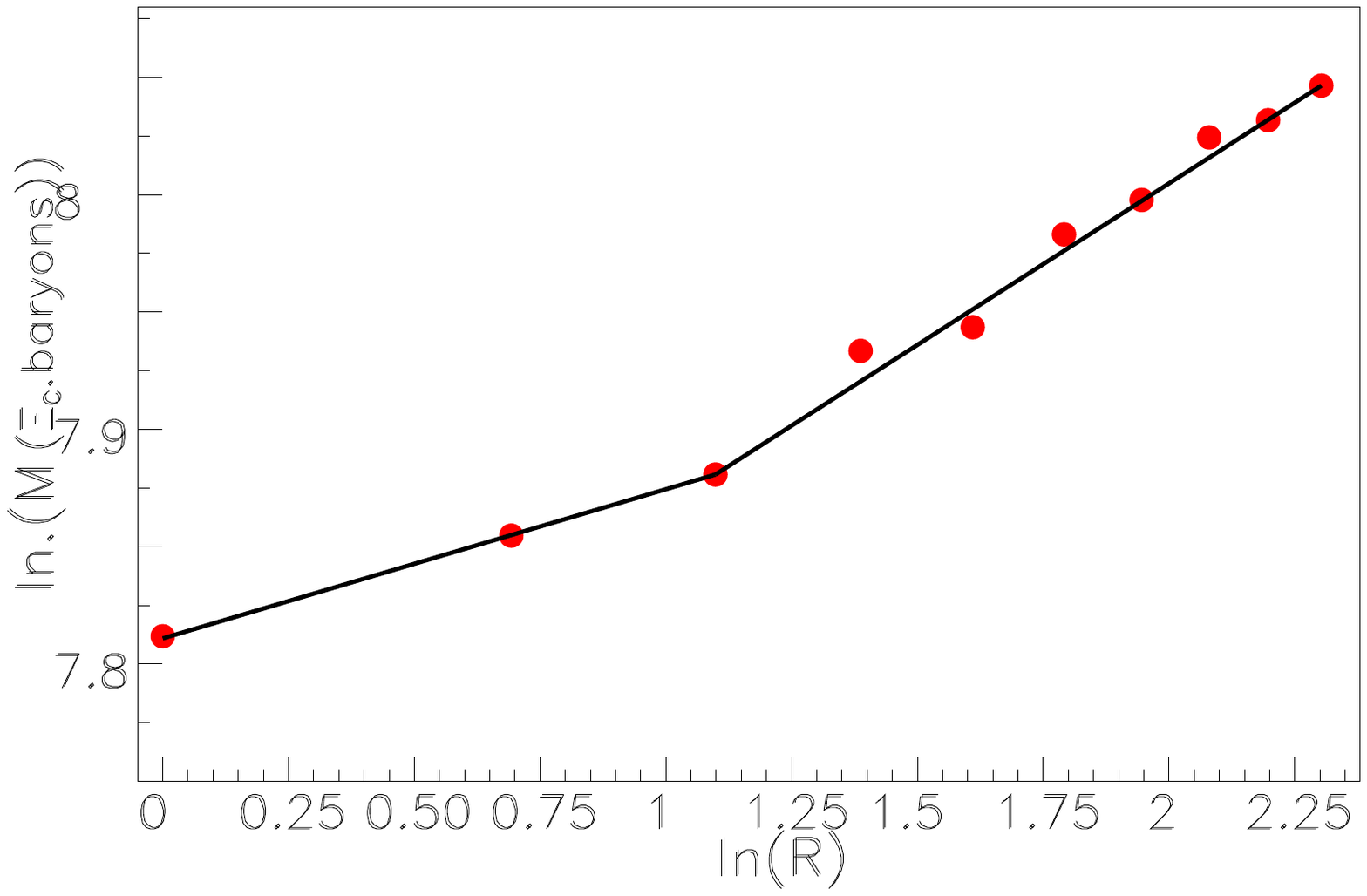}}
\end{center}
\end{figure}
 The ratio of $m_{r+1}/m_{r}$ masses, for the $\Xi_{c}$ baryons is shown in figure~28.
\begin{figure}[ht]
\begin{center}
\caption{Ratios of $m_{r+1}/m_{r}$ masses, for $\Xi_{c}$ baryons.}
\hspace*{-3.mm}
\includegraphics[bb=14 230 530 558,clip,scale=0.5]{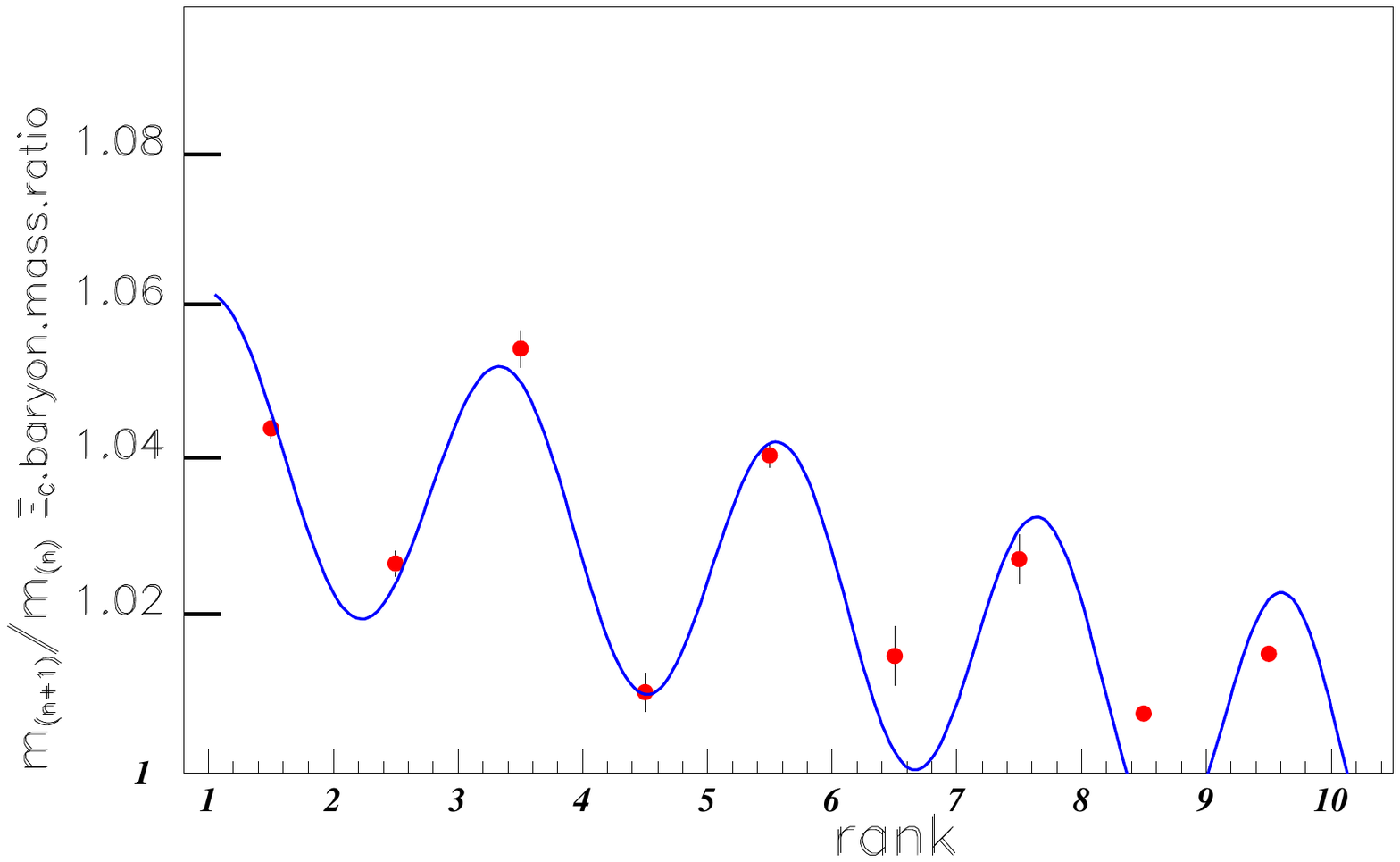}
\end{center}
\end{figure}
A nice fit is observed between the data and calculated distribution.

The heavier baryons are too scarce to allow the same discussion. Indeed PDG reports only two $\Omega_{c}$ masses, one $\Xi_{cc}$ mass, one $\Lambda_{b}$ mass, two $\Sigma_{b}$ masses, and one $\Xi_{b}$ mass.
\section{Discussion}
The hadronic mass ratios $m_{r+1}/m_{r}$ are correctly fitted over all distributions, in many species, with help of equation (3). This is only true for the first n$\approx$15 points, except for the first one. This is as well observed in the log-log representations, since the first point is not aligned with the followings, as in the $m_{r+1}/m_{r}$ mass ratio representations. These species are: unflavoured, strange, charmed, $c -{\bar c}$, and $b -{\bar b}$ mesons in the one side and N, $\Delta$, $\Lambda$, $\Sigma$, and $\Xi$ baryons in the other side. 
 
On the other hand, several other species do not exhibit such large ratio values at first "r" points, and are therefore correctly fitted over the total distribution. These species are: charmed strange mesons, charmed, and charmed-strange baryons, and also the exotic narrow mesons \cite{jy1,bt1,troyan1,troyan2,boris2}, exotic narrow baryons \cite{bor1,bor2,BT,bt2002} and exotic narrow  dibaryons \cite{filkov2001,11,12,btdibar} which were also analysed within fractal properties, but not illustrated here.

In order to check  the possibilty to observe fractal properties for the full widths $\Gamma$ of different hadron species, some log-log plots and the corresponding  $\Gamma_{r+1}/\Gamma_{r}$ ratios were studied. The total widths of most species are precise for the first masses, and then become quickly unprecise. Therefore only a relatively small number of species were kept for the study of the total width variation in the scope of fractal properties. This study is not presented here, in order to not lengthen too much the paper; it will be presented elsewhere. On the whole, the corresponding figures of the log-log plot of total widths show several straight lines, at least two straight lines, without superposition, suggesting a multiple fractal property. 

Since all $m_{r+1}/m_{r}$ mass ratio distributions are obtained with several adjustable parameters, it is important to study their variation from one species to another one. These variations are shown in the forthcoming figures showing the parameter values versus the masses. The mesons are shown by full circles (red on line), the baryons by full stars (blue on line), and the dibaryons marks by full triangles (green on line). The marks corresponding to narrow exotic species, are overmarked by empty squares. The horizontal lines, show the correctly fitted mass range of each species, using equarion (3). Each mark is plotted in the middle of each such horizontal line.  The precisions drawn on the parameters are arbitrary. In order only to guide the eye, the parameters of all distributions are joined by straight lines or smooth curves.

As already said, "$r_{c}$" is weakly defined, since the number of oscillations is not large. When "r" increases and comes close to "$r_{c}$", the oscillations contract, allowing to get the "$r_{c}$" value. This is observed only for the charmonium masses since in that species many precise masses exist. $r_{c}$ = 40 gives a good oscillation contraction for the charmonium distribution. We fix the same value $r_{c}$ = 40 for all species of our study, without attempt to move it.

We show, first, the distribution of the main fitted parameters describing the mass variations: the critical exponent "s", the parameter $a_{1}$ giving the amplitudes of the oscillations,  the fundamental scaling ratio $\lambda$~=~exp(1/$\Omega$). Then we show the calculated parameter "$\mu$" defined by the relation
$\mu$~=~$\lambda^{s}$.  "$\alpha$" which signs the presence of power laws and DSI, is given by Re~($\alpha$)~=~"-s"  and
Im~($\alpha$)~=~ 2i$\pi~\Omega$.

\begin{figure}[ht]
\begin{center}
\caption{Distribution of the fitted critical exponent "s" parameter, s = -Re ($\alpha$), from all hadronic species  (see text).}
\hspace*{-3.mm}
\scalebox{0.95}[1.3]{
\includegraphics[bb=2 270 530 520,clip,scale=0.5]{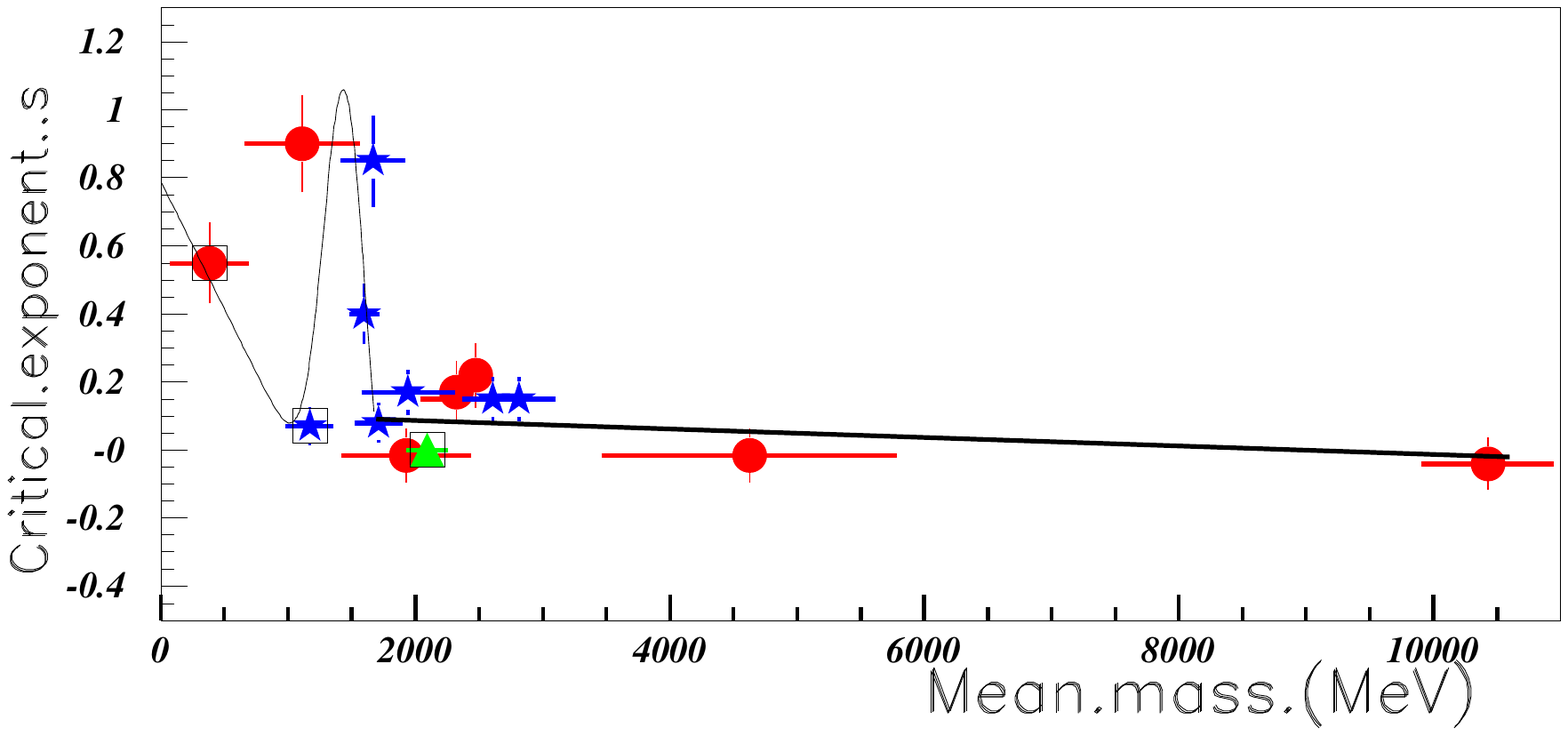}}
\end{center}
\end{figure}

\begin{figure}[ht]
\begin{center}
\caption{Distribution of the fitted $a_{1}$ parameter from all hadronic species  (see text).}
\hspace*{-3.mm}
\scalebox{0.95}[1.3]{
\includegraphics[bb=2 270 530 520,clip,scale=0.5]{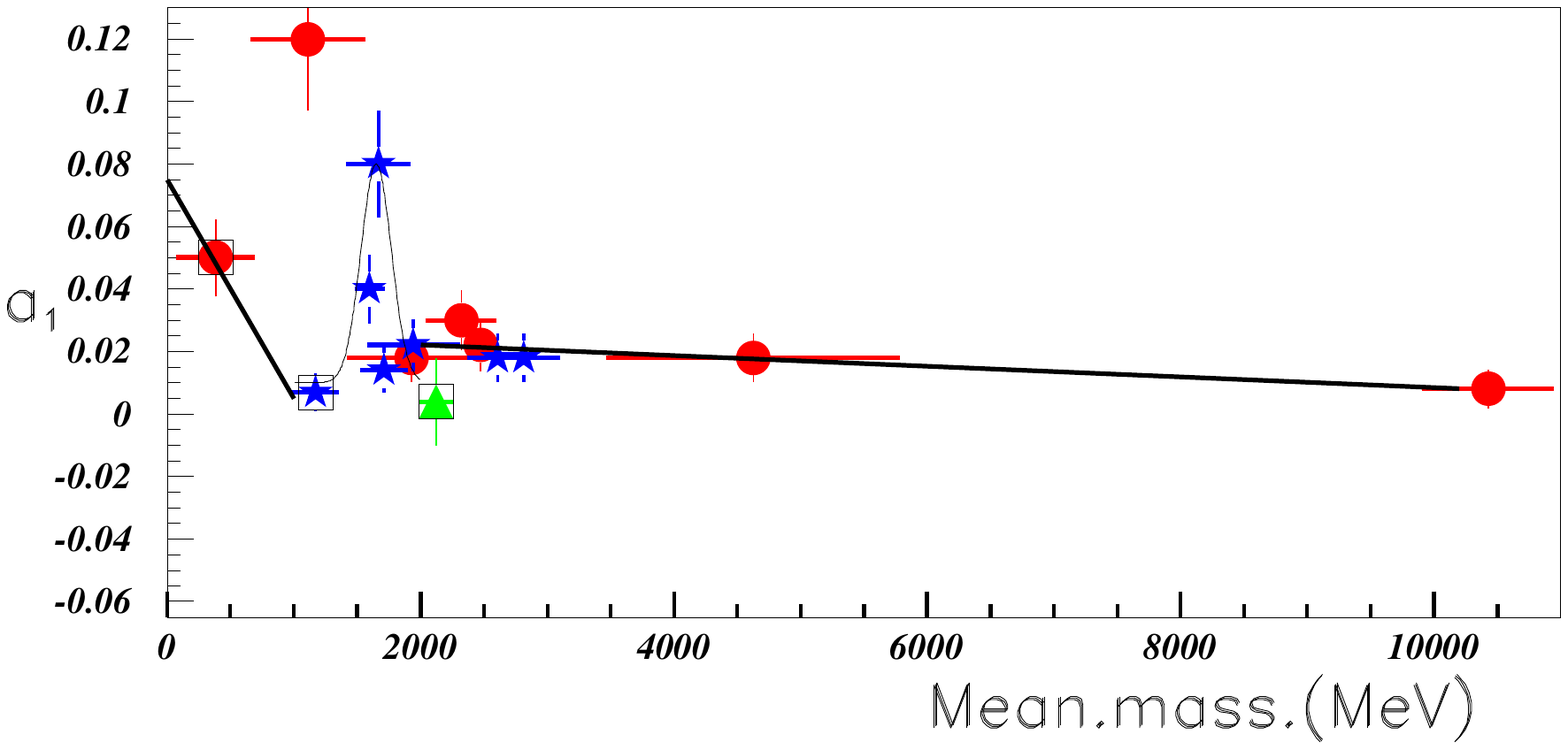}}
\end{center}
\end{figure}

\begin{figure}[ht]
\begin{center}
\caption{Distribution of the calculated parameter $\lambda$ from all hadronic species  (see text).}
\hspace*{-3.mm}
\scalebox{0.95}[1.3]{
\includegraphics[bb=2 270 530 520,clip,scale=0.5]{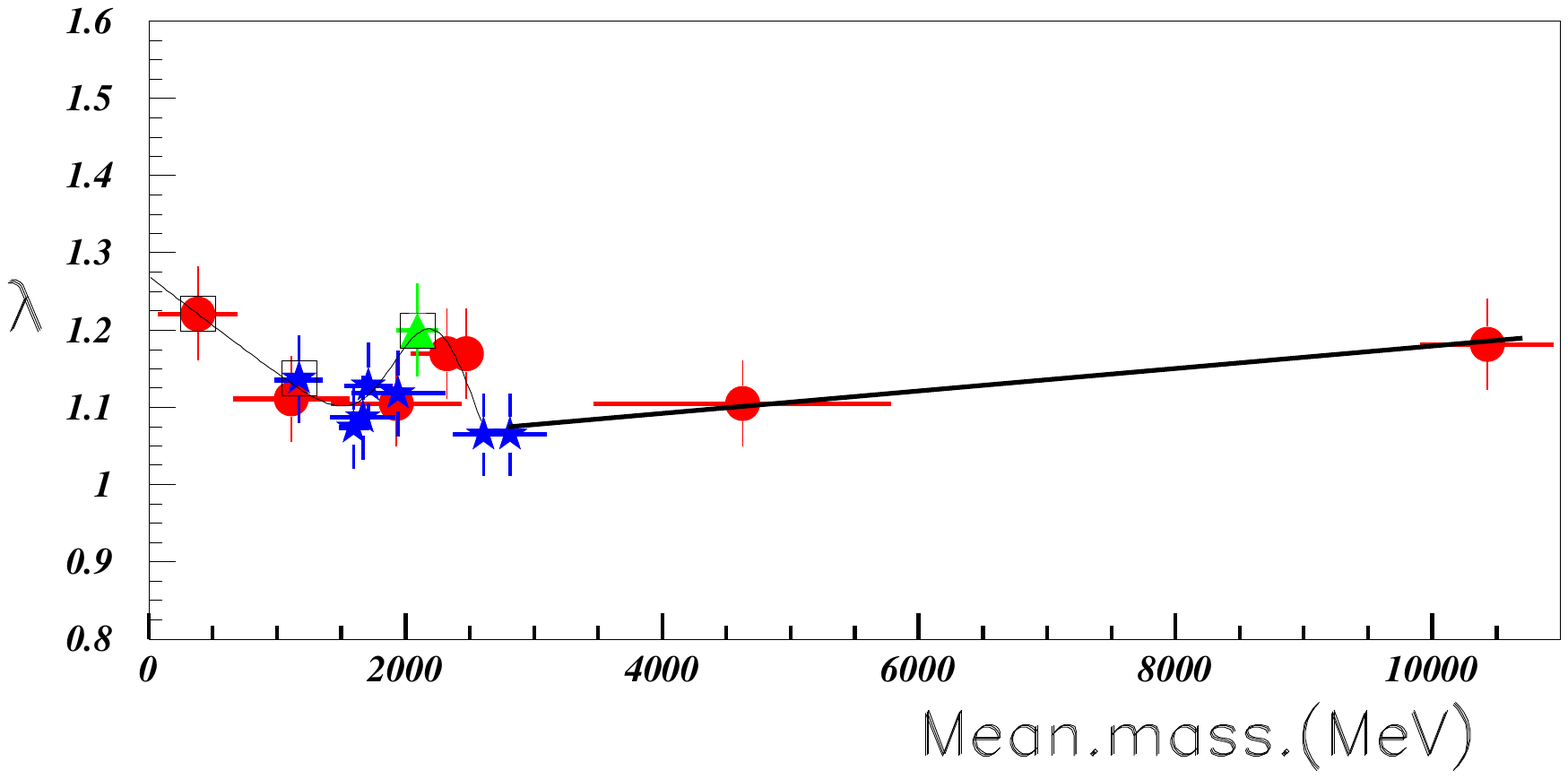}}
\end{center}
\end{figure} 

\begin{figure}[ht]
\begin{center}
\caption{Distribution of the calculated parameter $\mu$ from all hadronic species  (see text).}
\hspace*{-3.mm}
\scalebox{0.95}[1.3]{
\includegraphics[bb=2 270 530 520,clip,scale=0.5]{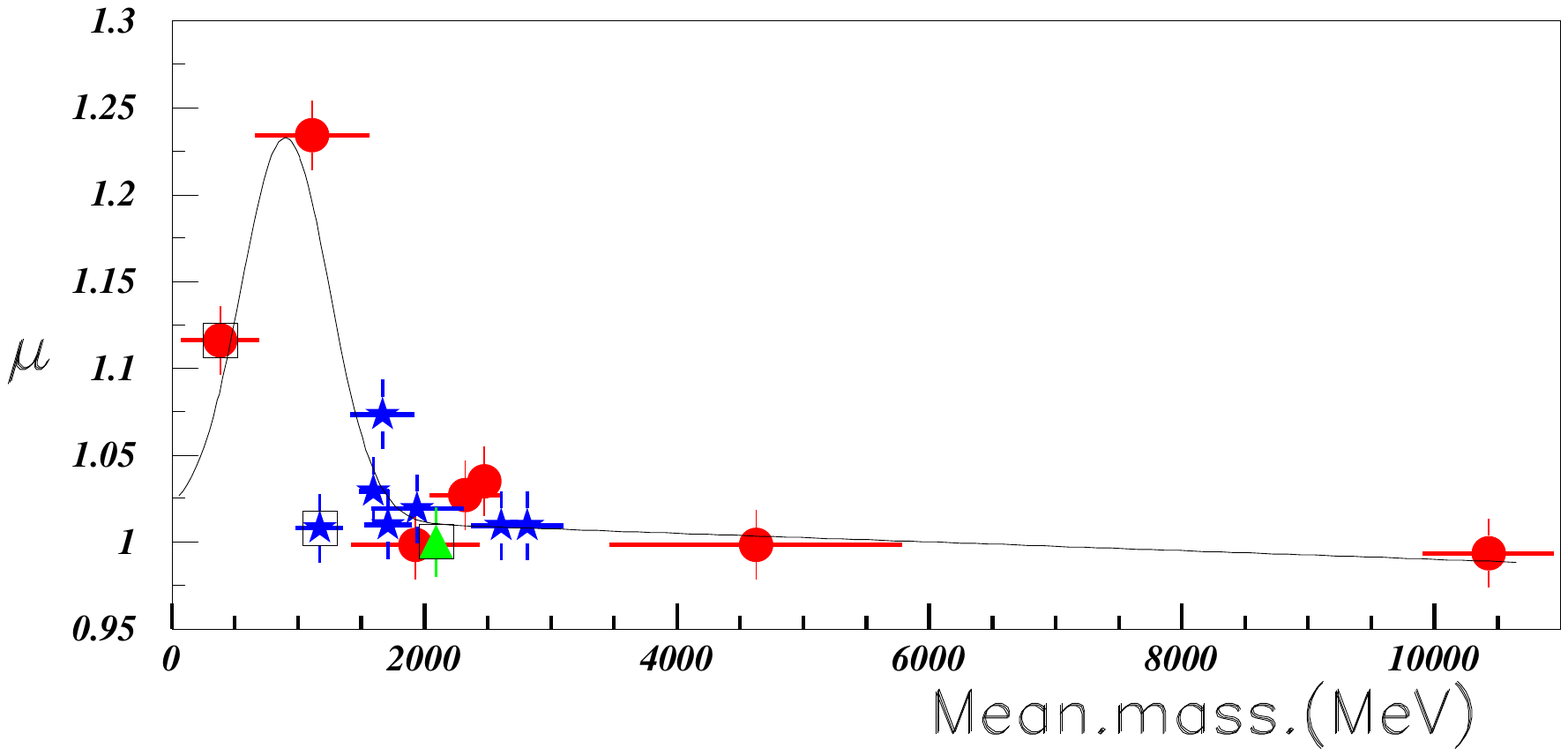}}
\end{center}
\end{figure} 

We observe also, that with a good precision, s is proportionnal to $a_1$.
The real part of $\alpha$ is very small, compared to the imaginary part, the ratio being generally smaller that 2*10$^{-2}$. The shapes of "s", "$\lambda$", "a$_{1}$", and "C" (not shown) distributions are approximately the same. In the same way, the shapes of $\Omega$, and $\Psi$ (not shown) are approximately the same.

The amplitudes of the log-periodic oscillations are given by the "a$_{1}$" parameter, which vary from 2~10$^{-2}$  up to 10$^{-1}$. This is not a small effect \cite{nottale}. 

Fig.~3 of \cite{sornette} reproduces the data  \cite{bernasconi} of a random walk process due to intermittent encounters with slow regions, and shows the corresponding fit to these data. It is noteworthy that both data and fit, look like the figures of hadronic mass ratios shown above. In these data \cite{bernasconi} also, the beginning of the distribution displays a high peak, not described by the fit.

We observe similar shapes between some $m_{r+1}/m_{r}$ mass ratio distributions. This concerns charmed mesons, charmed-strange mesons, and also although with a smaller extend, the charmonium mesons. Such observation suggests that the masses of these meson species should be compared, after global mass translation.
\begin{figure}[ht]
\begin{center}
\caption{Comparison of all PDG meson masses up to  M=3500~MeV, after global translations of each species, in order to equalize the first mass of each species to the first charm meson mass (see text). The figure is colored on line.}
\hspace*{-3.mm}
\scalebox{0.95}[1.3]{
\includegraphics[bb=8 30 525 525,clip,scale=0.5]{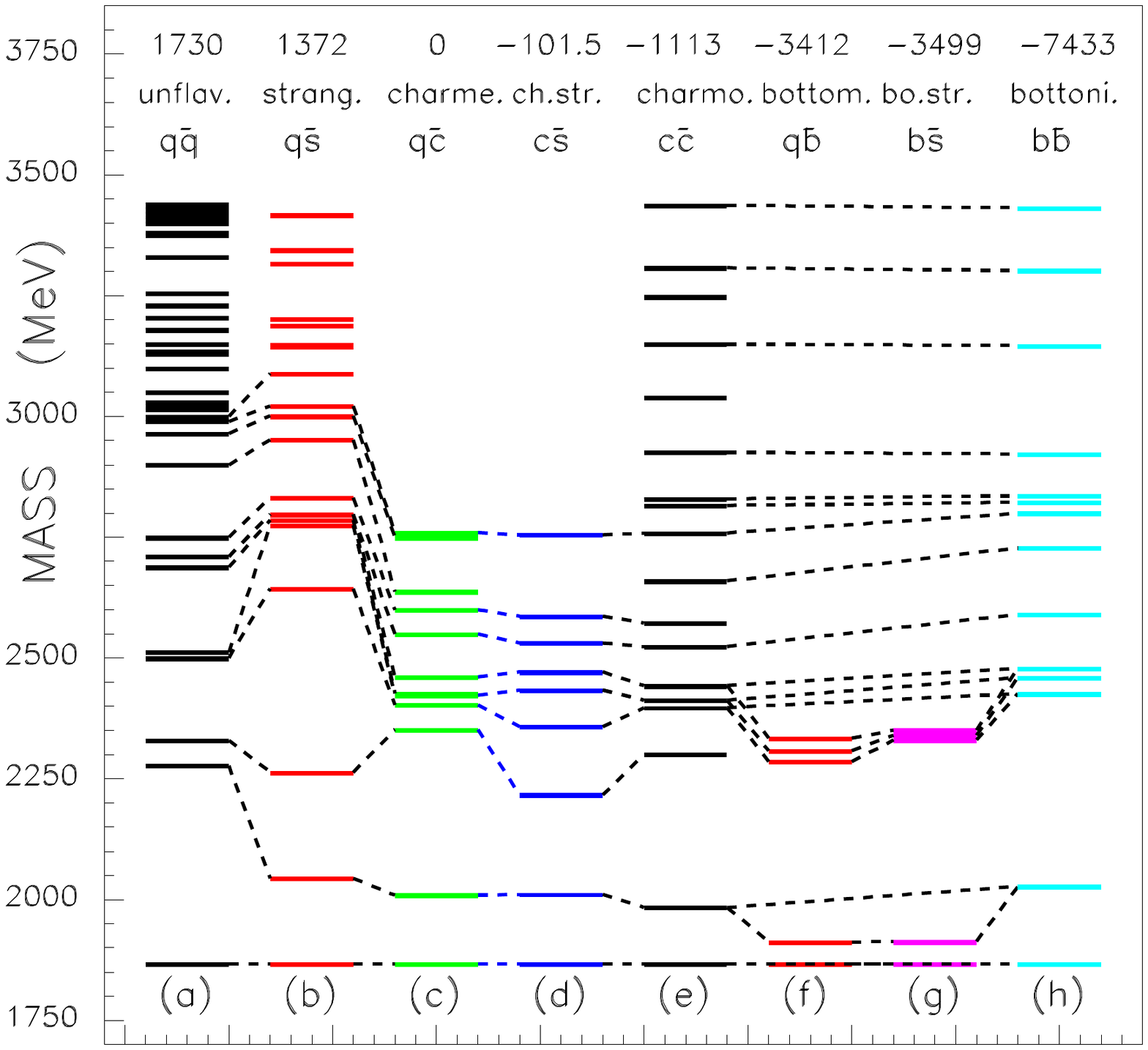}}
\end{center}
\end{figure} 
Figure~33 shows all PDG meson mass spectra up to M=3500~MeV, after global translations performed to equalize the fundamental mass of all species to the charmed meson fundamental mass: M=1867~MeV. The species are arranged by increasing fundamental masses. The correspondance between the columns and the given species is: 
(a):~unflavoured mesons, (b):~strange mesons, (c):~charmed mesons, (d):~strange-charmed mesons, (e):~charmonium mesons, (f):~bottom  mesons, (g): bottom-strange mesons, and (h): bottonium mesons. The quark contents are shown, as well as the amount of mass translation at the top of the figure. We observe indeed, after translation, very stable mass excitations between all three species containing a (two) charmed quark(s).

 We observe that the shapes of columns between unflavoured and strange mesons are different. However we observe similar masses (after translation) between charmonium and bottonium mesons (masses joined by dashed lines). Such observation allows us to tentatively predict the masses of some still unobserved bottonium mesons to be close to M$\approx$~9767~MeV, 10073~MeV, 10458~MeV, and 10662~MeV. 
In the same way, similar columns corresponding to charmed mesons, strange-charmed mesons, and charmonium mesons allows us to tentatively predict a still unobserved strange-charmed meson at at M$\approx$~2747~MeV.
\begin{figure}[ht]
\begin{center}
\caption{Comparison of all PDG baryon masses up to  M=2940~MeV, after global translations of each species, in order to equalize the first mass of all species to the first charmed baryon mass (see text). The figure is colored on line.}
\hspace*{-3.mm}
\scalebox{0.95}[1.3]{
\includegraphics[bb=8 30 525 525,clip,scale=0.5]{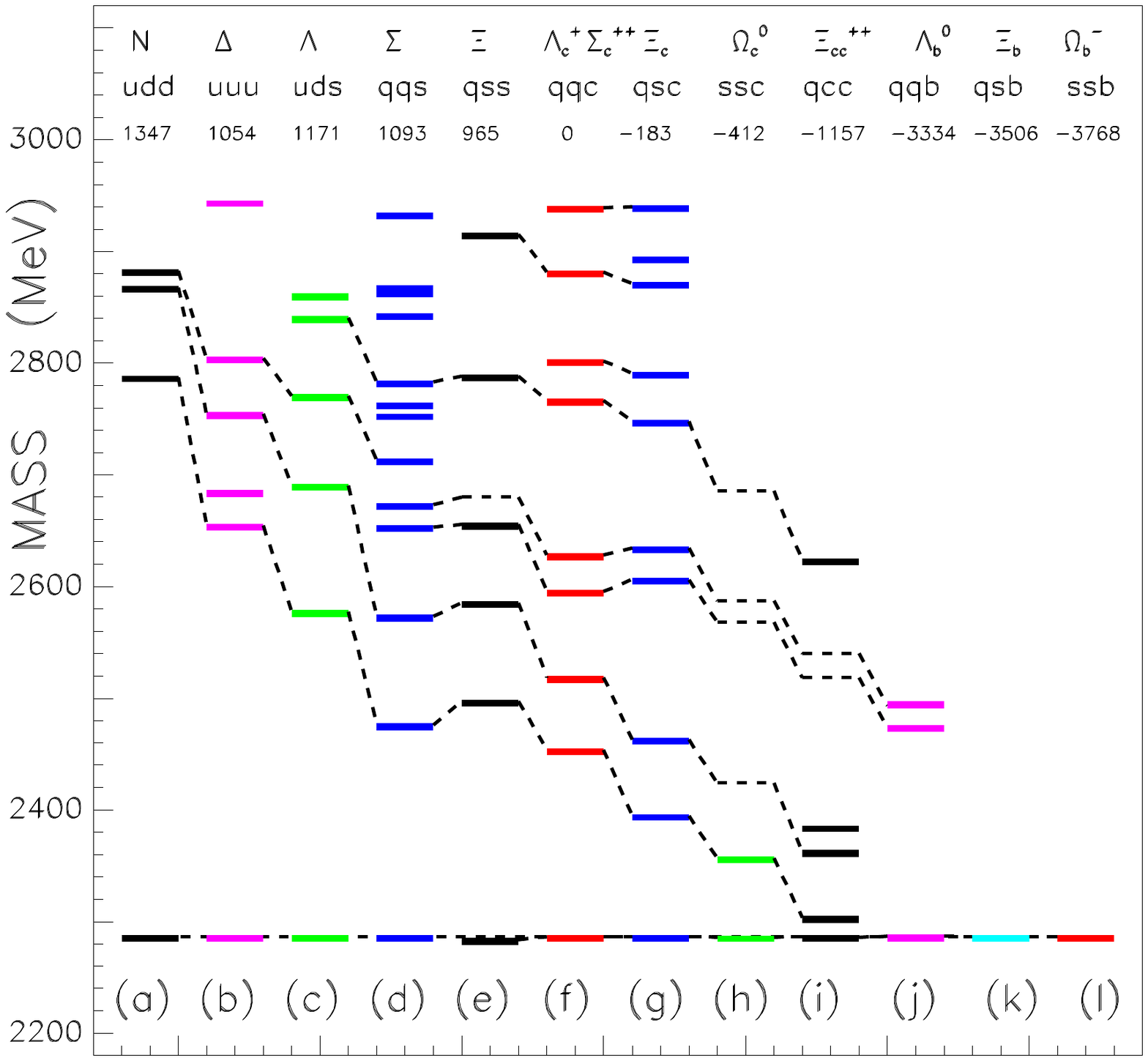}}
\end{center}
\end{figure} 

 Figure~34 shows a similar comparison between all PDG baryonic masses up to 2940~MeV, after a translation comparable as the one described above for mesons. Here the fundamental mass of all species, is ajusted to the mass of the charmed $\Lambda_{c}^{+}$  baryon. The species are arranged by increasing fundamental masses \cite{rem1}.
 We observe a regular mass decrease of the second and third masses of nearly all species, and also, although less regular, a mass decrease of the fourth and fifth masses.
The shift of the excited masses of all species, contract progressively when fundamental masses increase.  Using this regularity, several masses, still not observed, are tentatively predicted. They are shown in figure~40 by dashed lines. These masses are:

  M~$\approx$~1715~MeV,  for the $\Xi$ baryons,
 
   M~$\approx$~2836~MeV, 2980~MeV, 3000~MeV, and 3100~MeV for the $\Omega_{c}^{0}$ baryons,
  
   M~$\approx$~3675~MeV and 3697~MeV for the $\Xi_{cc}^{++}$ baryons.

The study of the spectroscopy of heavy bottom baryons will require rather good resolution.
\section{Conclusion}
We have shown that the masses of PDG display fractal properties and DSI. The straight lines in the log-log representations of the hadronic masses, namely the log of the studied quantity versus the log of the rank, show that they follow the fractal law.  In this model, complex critical exponents lead to log-periodic corrections to scaling. The ratio between adjacent hadronic masses exhibit clear oscillations in agreement with the cosine of the formula described in the log-periodic corrections of the DIS model. Nearly all masses of  mesonic and bosonic species, are well fitted by the equation (3) derived from the discrete-scale invariance model \cite{sornette}. Good quantitative fits are obtained except for the first oscillation sometimes well larger than obtained through formula, in agreement with the large slope in the log-log distributions between the first and the second points.

The fits are obtained with use of several parameters. However these
 parameters describing the successive mass ratios are not distributed randomly. Each parameter of a given species, is connected with the same parameter of all hadronic species through a continuous distribution. All these distributions show a structure in the region M~=~1500-2000~MeV. Then all successive mass ratios of all hadronic mesonic and baryonic species have a common connection, in spite of the large gap between 5 GeV and 10 GeV.

The same studies were done for the widths of some meson and baryon species. In spite of large uncertainties, we have observed quite systematically,
a multifractal property. Such observation deserves other theoretical study.

When the spectroscopy of the charmonium states is rich, the bottom counterparts of the higher baryonic masses has still to be observed. Moreover the actual knowledge of the $s{\bar s}$ mesons is still nearly unknown.

The nice fit observed for some hadronic species, allows us to predict possible masses of still unobserved hadrons. A few species are not studied, since a too small number of masses are presently observed. They are bottom strange and bottom charmed mesons.

Hybrid mesons are also outside the scope of the present study.

All figures shown in this paper, will be improved, when new mesonic masses will be extracted from experiemnts, or when some omitted from summary table masses,  will be definitively removed. These masses are introduced tentatively in our study. These modifications should in general concern "large" masses, therefore "large" rank.

The fractal property of the hadronic masses, increases the many fractal aspects observed in the universe.
The agreement with the theoretical relation (3) suggests a possible new physical property of hadronic masses.

In conclusion this work shows that the hadronic masses obey to the log-periodic fractal model and DSI.
\section{Acknowledgments}
Ivan Brissaud introduces me to the study of the hadronic masses inside fractals. I thank him for his stimulating remarks and interest.

\vspace*{0.4cm}
\section{Appendix}
\begin{table}[h]
\begin{center}
\caption{Mesonic masses used in previous figures.}
\label{Table I}
\begin{tabular}[t]{c c c }
\hline
species&figs.&masses (in MeV)\\
\hline
Mesons\\
Unflav.&1-3&137, 547.85, 600, 775.49, 782.65\\
&&957.78, 980, 980, 1019.455, 1170\\
&&1229.5, 1230, 1275.1, 1281.8, 1294\\
&&1300, 1318.3, 1370, 1354, 1386\\
&&1409.8, 1425, 1426.4, 1430, 1465\\
&&1474, 1476, 1505, 1518, 1525, 1562\\
&&1570, 1594, 1617, 1639, 1647, 1662\\
&&1667, 1672.4, 1680, 1688.8, 1720\\
&&1720, 1732, 1815, 1816, 1833.7\\
&&1842, 1854, 1895, 1900, 1903, 1944\\
&&1982, 1990, 2011, 2090, 2103, 2149\\
&&2157, 2175 2189, 2231.1, 2226, 2250\\
&&2297, 2300, 2330, 2330, 2339, 2340\\
&&2450, 2468\\
Strange&4, 5&495, 672, 891.66, 1272, 1403, 1414,\\
&&1425, 1425.6, 1460, 1580, 1629, 1650\\
&&1717, 1773, 1776, 1816, 1830, 1945\\
&&1973, 2045, 2247, 2324, 2382, 2490\\
&&3054\\
Charmed&6, 7&1867, 2009, 2318, 2403, 2423, 2427\\
&&2461.1,2539.4 2608.7 2637 2710\\
&&2752.4, 2763.3, 2860\\
Charm.-stran.&8, 9& 1968.47, 2112.3, 2317.8, 2459.5\\
&&2535.29,2572.6, 2632.6, 2688, 2709\\
&&2862, 3044\\
Bottom&10&5279.17, 5325.1, 5698, 5723.4, 5743\\
Botto.-Stran.&11&5366.3, 5415.4, 5829.4, 5839.7, 5850\\
$c-{\bar c}$&12, 13& 2980.3, 3096.916, 3414.75, 3510.66\\
&&3525.93, 3556.2, 3637, 3686.09\\
&&3772.92, 3872.2, 3915.5, 3929, 3943\\
&&4039, 4143, 4153, 4156, 4248, 4263\\
&&4350, 4361, 4421, 4443, 4550, 4664\\
&&4780, 4870, 5090, 5300, 5440, 5660\\
&& 5910\\
$b-{\bar b}$&14, 15&9300, 9460.3, 9859.44, 9892.78\\
&&9912.21, 10023.26, 10161.1, 10232.5\\
&&10255.46, 10268.65, 10355.2\\
&&10579.4, 10735, 10865, 11019\\
\hline
\end{tabular}
\end{center}
\end{table}
\samepage
\begin{table}[h]
\begin{center}
\caption{Baryonic masses used in previous figures.}
\label{Table II}
\begin{tabular}[t]{c c c}
\hline
species&figs.&masses  (in MeV)\\
\hline
Baryons\\
N&16, 17&939, 1440, 1520, 1535, 1655, 1675, 1685\\
&&1700, 1710, 1720, 1900, 1990, 2000, 2080\\
&&2090, 2100, 2190, 2200, 2220, 2275, 2600\\
&&2700, 3000\\
$\Delta$&18&1232, 1600, 1630, 1700, 1750, 1890, 1900\\
&&1910, 1920, 1930, 1940, 1940, 2000, 2150\\
&&2200, 2300, 2350, 2390, 2400, 2420, 2750\\
&&2950\\
$\Lambda$&19, 20&1115.68, 1405, 1520, 1600, 1670, 1690\\
&&1800, 1810, 1820, 1830, 1890, 2000, 2020\\
&&2100, 2110, 2325, 2350, 2585\\
$\Sigma$&21, 22&1193, 1385, 1480, 1560, 1580, 1620\\
&&1660, 1670, 1690, 1750, 1770, 1775, 1840\\
&&1880, 1915, 1940, 2000, 2030, 2070, 2080\\
&&2100, 2250, 2455, 2620, 3000, 3170\\
$\Xi$&23, 24& 1318.285, 1530, 1620, 1690, 1820, 1950\\
&&2030, 2120, 2250, 2370, 2500\\
Charmed&25, 26&2286.46, 2453.76, 2518.4, 2595.4\\
&&2628.1, 2766.6, 2802, 2881.53, 2939.3\\
$\Xi_{C}$&27, 28&2469.5, 2577.8, 2645.8, 2789.2, 2817.4\\
&&2931, 2974, 3054.2, 3077, 3122.9\\
\hline
\end{tabular}
\end{center}
\end{table}


\begin{thebibliography}{99}
\bibitem{mandelbrot}B. Mandelbrot, {\it Les objets fractals} (Flammarion, Paris 1975), {\it ibid} {\it The Fractal geometry of Nature} (Freeman, San Francisco, 1982). 
\bibitem{nottale1}L. Nottale, The Theory of Scale Relativity: Non-Differentiable Geometry and Fractal Space-Time. In : Computing Anticipatory Systems. CASY'S03 - Sixth international Conference (Li\`ege 2003), D.M. Dubois, Ed., American Institute of Physics Conference proceedings, 718, p.68 (2004).
\bibitem{btib}B. Tatischeff and I. Brissaud, arXiv:1005.0238v1 [hep-ph] (2010).
\bibitem{boris}B. Tatischeff, arXiv:1104.5379v1 [physics.gen-ph] (2011).
\bibitem{pascalutsa}V. Pascalutsa, Eur. Phys. J. A {\bf 16}, 149 (2003). 
\bibitem{goldfain}E. Goldfain, Electronic Journal of Theoretical Physics, 
{\bf 7, $n^{0}$ 23}, 75 (2010).
\bibitem{brodsky}S.J. Brodsky, Eur. Phys. J A {\bf 31}, 638 (2007).
\bibitem{sornette}D. Sornette, Physics Reports {\bf 297}, 239 (1998).
\bibitem{nottale}J. Chaline, L.Nottale, and P. Grou, avec la participation d'Ivan Brissaud; "Des Fleurs pour Schr$\ddot{o}$dinger, la Relativit\'e d'\'echelle et ses applications", ed. Ellipses \'Editions, 2009.
\bibitem{pdg}K. Nakamura and Particle Data Group, J. Phys. G: Nucl. Part. {\bf 37}, 075021 (2010).
\bibitem{anisovitch}A.V. Anisovitch {\it et al.}, Phys. Lett. B {\bf 491}, 47 (2000); {\it ibid} Phys. Lett. B {\bf 517}, 261 (2001); {\it ibid}  Phys. Lett. B {\bf 542}, 8 (2002); {\it ibid}  Phys. Lett. B {\bf 542}, 19(2002).
\bibitem{bugg}D.V. Bugg, Phys. Rep. {\bf 397}, 257 (2004).
\bibitem{babar}J. Benitez {\it et al.} (BaBar Collaboration), ICHEP2010, (2010);
P. del Amo Sanchez {\it et al.}, arXiv:1009.2076v1 [hep-ex] (2010).
\bibitem{babar1}B. Aubert {\it et al.} (BaBar Collaboration), Phys. Rev. D {\bf 80}, 092003 (2009); P. del Amo Sanchez {\it et al.} (BaBar Collaboration), arXiv:1009.2076 [hep-ex].
\bibitem{zhong}Xian-Hui Zhong, arXiv:1009.0359v1 [hep-ph] (2010).
\bibitem{wang}Zhi-Gang Wang, arXiv:1009.3605v1 [hep-ph] (2010).
\bibitem{li}De-Min Li, Peng-Fei Ji, and Bing Ma, Eur. Phys. J. C.  {\bf 71}, 1582 (2011).
\bibitem{aubert}B. Aubert  {\it et al.}, Phys. Rev D {\bf 74}, 091103(R) (2006).
\bibitem{ablikim}M. Ablikim {\it et al.} (BES Collaboration),  Phys. Rev. Lett. {\bf 100}, 102003 (2008).
\bibitem{hua}H.-X. Chen, A. Hosaka, and S.-L. Zhu, arXiv:0806.1998v1 [hep-ph] (2008).
\bibitem{olsen}S.L. Olsen, Nucl. Phys. {\bf A 827}, 53c (2009).
\bibitem{barkov}B.P. Barkov {\it et al.}, JETP Letters {\bf 70}, 248 (1999).
\bibitem{selex}A.V. Evdokimov {\it et al.}, arXiv:0406045 [hep-ex] (2004).
\bibitem{santoro}V. Santoro on behalf of the BABAR Collaboration, Nucl. Phys. B (Proc. Suppl.) {\bf 187}, 175 (2009).
\bibitem{br1}E. van Beveren and G. Rupp, arXiv:hep-ph/0605317.
\bibitem{br2}E. van Beveren and G. Rupp, Phys. Rev D {\bf 80}, 074001 (2009).
\bibitem{br3}E. van Beveren and G. Rupp, arXiv:1004.4368v1 [hep-ph] (2010).
\bibitem{br4}E. van Beveren and G. Rupp, arXiv:0910.0967 [hep-ph] (2009).
\bibitem{giurgiu}G. Giurgiu on behalf of the CDF Collaboration,  Nucl. Phys. B (Proc. Suppl.) {\bf 187}, 44 (2009).
\bibitem{lipkin}H.J. Lipkin, Nucl. Phys. A {\bf 675}, 443c (2000), S.U. Chung, Nucl. Phys. A {\bf 675}, 453c (2000),
\bibitem{guo}Feng-Kun Guo, C. Hanhart, and Ulf-G. Meissner, Phys. Lett. B {\bf 665}, 26 (2008).
\bibitem{aubert9}B. Aubert {\it et al.}, Phys. Rev D {\bf 77}, 111101(R) (2008).
\bibitem{ablikim9}M. Ablikim {\it et al.}, Phys. Rev D {\bf 70}, 012003 (2004).
\bibitem{abey}K. Abe {\it et al.}, arXiv:0708.1790 [hep-ex] (2007).
\bibitem{ke}Hong-Wei Ke and Xiang Liu, arXiv:0806.0998v2 [hep-ph] (2008).
\bibitem{jy1}J. Yonnet {\it et al.}, Phys. Rev. C{\bf 63}, 014001 (2000).
\bibitem{bt1}B. Tatischeff {\it et al.}, Phys. Rev. C{\bf 62}, 054001 (2000).
\bibitem{troyan1}Yu.A. Troyan {\it et al.}, Proceedings of the XVI International Baldin Seminar on High Energy Physics Problems, Dubna, p. 163 (2002).
\bibitem{troyan2}Yu.A. Troyan {\it et al.}, JINR Rapid Communications {\bf 6} 80, p73 (1996).
\bibitem{boris2}B. Tatischeff and E. Tomasi-Gustafsson, Phys. of Part. and Nucl.
 Lett. {\bf 5}, 363 (2008); {\it ibid}  Phys. of Part. and Nucl.
 Lett. {\bf 5}, 420 (2008).
\bibitem{bor1}B. Tatischeff {\it et al.}, Phys. Rev. Letters {\bf 79}, 601 (1997).
\bibitem{bor2}B. Tatischeff {\it et al.}, Eur. Phys. J. A {\bf 17}, 245 (2003).
\bibitem{BT}B. Tatischeff {\it et al.}, Phys. Rev. C {\bf 72}, 034004 (2005).
\bibitem{bt2002}B. Tatischeff Proc. XVI Inter. Baldin Sem. on High Energy Phys. Problems, p. 153 (2002); {\it ibid}arXiv:nucl-ex/0207004 (2002).
\bibitem{filkov2001}L. V. Filkov {\it et al.}, Eur. Phys. J {\bf A 12},369 (2001). 
\bibitem{11}B. Tatischeff  {\it et al.}, Phys. Rev. C{\bf 36}, 1995 (1987).
\bibitem{12}B. Tatischeff  {\it et al.}, Europhys. Lett. {\bf 4}, 671 (1987); Z. Phys. A
{\bf 328}, 147 (1987).
\bibitem{btdibar}B. Tatischeff  {\it et al.}, Phys. Rev. C{\bf 59}, 1878 (1999).
\bibitem{bernasconi}J. Bernasconi and W.R. Schneider, J. Phys. A {\bf 15}, L729 (1983).
\bibitem{rem1}The absence of spin and isospin in this study, may explain the inversion of $\Delta$ and $\Lambda$ columns.  
\end{thebibliography}
\end{document}